\documentstyle[eqsecnum,aps,epsfig]{revtex}

\begin{document}
\title{
Balloon Measurements of Cosmic Ray Muon Spectra in the Atmosphere 
along with those of Primary Protons and Helium Nuclei over Mid-Latitude        
}
\author{R. Bellotti, F. Cafagna, M. 
Circella\thanks{Electronic address: circella@ba.infn.it} and   C. N. De 
Marzo }
\address{Dipartimento di Fisica dell'Universit\`a di Bari and 
INFN-Sezione di Bari,\\
Via Amendola 173, 70126 Bari, Italy} 
\author{
R. L. Golden\thanks{Deceased.} and S. J. Stochaj } 
\address{
Particle Astrophysics Laboratory, 
New Mexico State University,\\
 Las Cruces, New Mexico 88003}

\author{M. P. De Pascale, A. Morselli and
P. Picozza}
\address{Dipartimento di Fisica dell'Universit\`a di Roma
 ``Tor Vergata'' and INFN-Sezione di Roma II, \\
Via Carnevale 15, 00173 Rome, Italy}

\author{ S. A. Stephens}
\address{Tata Institute of Fundamental 
Research, Bombay 400 005, India.}

\author{ M. Hof, W. Menn and M. Simon} 
\address{Universit\"at Siegen, 
57068 Siegen, Germany}

\author{J. W. Mitchell,
J. F. Ormes and R. E. Streitmatter }
\address{NASA Goddard Space Flight Center, Greenbelt, Maryland 20771}
\author{
N. Finetti, C. Grimani\thanks{Also at Dipartimento di Fisica
dell'Universit\`a di Urbino, Urbino, Italy.}, 
P. Papini\thanks{Electronic address: papini@fi.infn.it }, S. Piccardi and
P. Spillantini }
\address{Dipartimento di Fisica dell'Universit\`a di Firenze and
INFN-Sezione di Firenze,\\
Largo Enrico Fermi 2, 50125 Florence, Italy} 
\author{
G. Basini and M. Ricci }
\address{
INFN Laboratori Nazionali di Frascati, 
Via Enrico Fermi 40, 00044 Frascati, Italy}
\date{March 6, 1999) \\
(To appear in Phys.\ Rev.\ D}
\maketitle
\begin{abstract}
We report here  the measurements of the energy spectra of 
atmospheric muons and of  
the parent cosmic ray primary proton and helium 
nuclei in a single experiment. These were carried out using the 
MASS superconducting spectrometer in a balloon flight 
experiment in 1991.
The relevance of these results to the atmospheric neutrino anomaly
is emphasized. In particular, 
this approach allows 
uncertainties caused by 
the level of solar modulation, 
the geomagnetic cut-off of the primaries and  possible 
experimental systematics to be decoupled 
in the comparison of calculated fluxes of
muons to measured muon fluxes.
The muon observations cover the momentum and depth ranges 
of 0.3--40 GeV/$c$
and 5--886 g/cm$^2$, respectively. A comparison of these 
results with those obtained
in a previous experiment by the same 
collaboration using a similar 
apparatus allows us to search for differences due to the 
different experimental conditions at low energy and to check 
for the overall normalization between the two measurements. The proton 
and helium primary measurements cover the rigidity range from 
3 to 100 GV, in which both the solar modulation 
and the geomagnetic cut-off 
affect the energy spectra at low energies.
From the observed low-energy helium spectrum,
the 
geomagnetic transmission function at mid-latitude has been
determined.
\end{abstract}
\pacs{PACS numbers: 96.40.-z, 96.40.De, 96.40.Kk, 96.40.Tv, 14.60.P, 14.60.S}
\newpage
\section{ Introduction}
\begin{figure}
\mbox{\epsfig{file=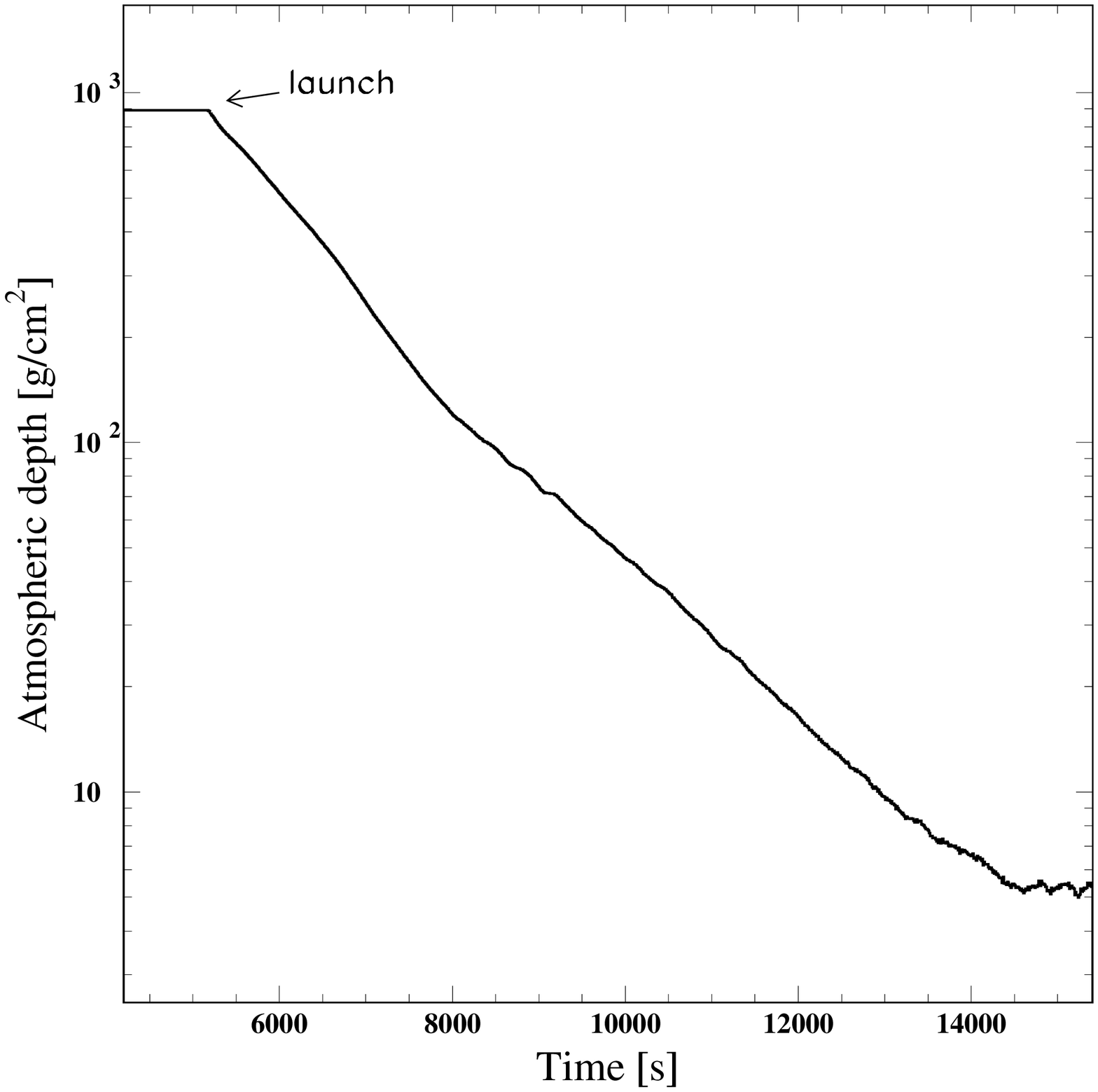,width=16cm}}
\caption{Ascent curve of the payload, based on the pressure
measurements. 
Time is measured from the startup of the on-board computer and 
the launch is at 5180 s from this reference time.
\label{fascent}}
\end{figure}

Data on muon spectrum as a function of the atmospheric depth
in the momentum interval 0.3--40~GeV/$c$ have been published 
earlier by this collaboration~\cite{PRD}. We report 
in this paper a new measurement of muon spectra in 
the atmosphere  
as well as the spectra of proton and helium nuclei which were 
measured at the float altitude with the same apparatus during the same 
balloon flight. The measurements were performed with the 
MASS (Matter Antimatter Spectrometer System) apparatus 
on September 23, 1991 starting from Ft. Sumner, NM at 1270 m above 
sea-level. The coordinates of this location 
are~$34^{\circ}$N and~$104^{\circ}$W, corresponding to an effective 
vertical cut-off rigidity of about~4.3~GV. 
The balloon ascent lasted for almost 3 hours, during which about 
240,000 triggers were collected. 
The muon measurements cover the altitude range 
from ground level to 
36~km, which corresponds to 
about 5 g/cm$^2$ of atmospheric depth. 
The ascent curve of the apparatus, 
based on the pressure measurements taken by the 
payload sensors 
is shown in Fig.~\ref{fascent}. 
The float data analyzed for this work cover an exposure time of
about 10 hours. These data were taken at atmospheric depths between 
4 and 7 g/cm$^2$, with an average value of 5.8~g/cm$^2$.

Primary cosmic ray particles, while entering the Earth 
atmosphere, interact with the atmospheric nuclei and produce secondary 
particles (see \cite{Gaisserbook}, for an excellent introduction). 
Among the primary cosmic rays, protons and helium nuclei are the 
major components, and as a consequence, a large fraction of these secondary 
particles are produced by them. Most of the secondary particles decay and 
some of the decay products are muons and neutrinos. Muons and muon neutrinos 
are the decay products of mesons, and both muon and electron neutrinos are 
the result of muon decays.
Both these kinds of neutrinos are detected by underground detectors. \\
Due to this close relationship, atmospheric muons have been
often considered as a powerful tool to calibrate the 
calculations of atmospheric propagation, 
in particular for the neutrino flux evaluation 
(e.g., \cite{Stanevven,Perkinscal}).
This situation appears to be most interesting in 
the 
context of 
the increasing evidence of the atmospheric neutrino anomaly (for a 
recent discussion, see~\cite{Stanev}). The anomaly 
is based on the discrepancy between the observed ratio of 
the number of neutrino interactions due to $\mu$- type and that due 
to $e$-type, as measured by some underground detectors, 
\cite{KAMIOKA,IMB,SOUDAN,FREJUS,NUSEX,SUPERK}
and the robust theoretical expectation at low energy.
While the evidence for the anomaly will not be discussed here,
it is important to note that any interpretation of the phenomenon
depends crucially on the absolute value of the 
expected fluxes of neutrinos. 
\\
In order to take into account the details of particle propagation 
and interactions in the calculations of atmospheric cascades,
both analytic  \cite{Stephenspion,Bugaev,Liparical}
and Monte Carlo approaches~\cite{Barr,Honda,Lee} have been 
succesfully undertaken in the past.
An extensive work has investigated the differences 
between the recent neutrino calculations \cite{Gaissercomp},
indicating that the parametrization of the cross-sections for
meson creation in proton collisions with the 
atmospheric nuclei is one of 
the major reasons for this
discrepancy. 
It is well known that at low transverse momentum
$p_T$ the perturbative quark model 
does not work and, moreover, the data available from accelerator 
measurements are not enough to discriminate between different 
interaction models in the central collision region (Feynman 
$x_{L}\leq$0.1). 
Contribution from this experimentally unexplored region is 
important for the meson production.
An additional factor of inaccuracy may come from
the kinematics of the particle propagation and decay. Although
these processes are well known, their description in the atmospheric 
simulation codes requires some approximations.
In fact, most of the calculations published so far
are performed under the approximation of unidimensional propagation of 
the secondaries, and the effect of  this approximation 
on the low-energy neutrino results is still under study. \\
Another important input to the atmospheric propagation calculations, 
which may introduce a further degree of uncertainty, 
is the primary cosmic ray composition and flux.
The direct measurements of the primary components 
show sometimes significant discrepancies 
with respect to one another (see \cite{Papiniprot},
for a compilation). 
The differences in the experimental results 
are to some extent due to the specific 
conditions of the measurements, namely, the geomagnetic suppression 
and the solar modulation, and in part may be due
to experimental inaccuracies. 
Both the geomagnetic and solar cycle effects on the 
primary cosmic rays need to be taken into account to evaluate
the neutrino fluxes, since the 
undergound experiments 
collect events coming from a large interval
of geomagnetic locations over significant fractions of the 
solar activity cycles. 
While the geomagnetic suppression is a well understood mechanism
and significant improvements in its description have been 
introduced recently \cite{Liparigeom}, 
the solar modulation of cosmic rays 
is not exactly periodic and shows some pecularities (e.g., the
so-called ``Forbush events'') that are hard to describe in a model.
\\
A comparison of the expected muon fluxes to measurements of muons in the
atmosphere may help in reducing
the uncertainty in the neutrino calculations due to 
the above 
factors, namely the primary spectra and the interaction cross-sections; 
both affect to similar extent the muon and neutrino flux calculations. 
An obvious limitation to this approach is that the muon measurements are 
not always available in experiments, by which primary particle 
spectra are measured, and calculations are carried out using available 
primary spectra measured at a time and location, which may not correspond 
to the muon measurements.
The approach described in this investigation to measure the 
primary spectra of protons and helium nuclei along with the 
measurement of atmospheric muons by the same experiment, allows the 
following possibilities.
(i) The measured primary spectra can be used as input to the 
propagation calculations whose results have to be compared to the 
muon measurements, thus taking automatically into account the 
specific levels of geomagnetic suppression and solar modulation of 
the experiment. (ii) Possible systematics on the global normalization 
of the experiment (e.g., geometric factor, acquisition efficiency, 
etc.) will be compensated as well in such calculations. \\
While muon measurements at sea-level are widely reported in the 
literature, there have been very few attempts to measure the muon 
flux as a function of altitude. The early
experiments were performed either with airplane-borne apparatus 
or at mountain sites \cite{Conversi,Blokh}. Counter telescopes 
were used for detecting charged particles and muons
were usually selected by requiring them to traverse large 
amounts of matter without interacting.
The main difficulty in such experiments 
was to properly identify muons 
while rejecting the other components of the ``hard'' radiation. 
This problem was of course more
complex for positive muon measurements, 
since the proton flux rapidly increases with increasing altitude.
A thorough review of these earlier results is presented in~\cite{Tesi}.
The deployment of balloon-borne detectors allows 
the investigation to be extended to
momentum and depth ranges much larger than in previous experiments
\cite{PRD,IMAX,HEAT95,HEAT97}. \\
Preliminary results for the muon measurements from this study
were reported earlier
\cite{Roma,Durban},
as well as  preliminary proton results 
at float level \cite{papini2}.  
The measurement of the muon flux and charge ratio at the float level 
from this experiment has already been published \cite{Codino}.
\section{Detector Setup} 
\begin{figure}
\mbox{\epsfig{file=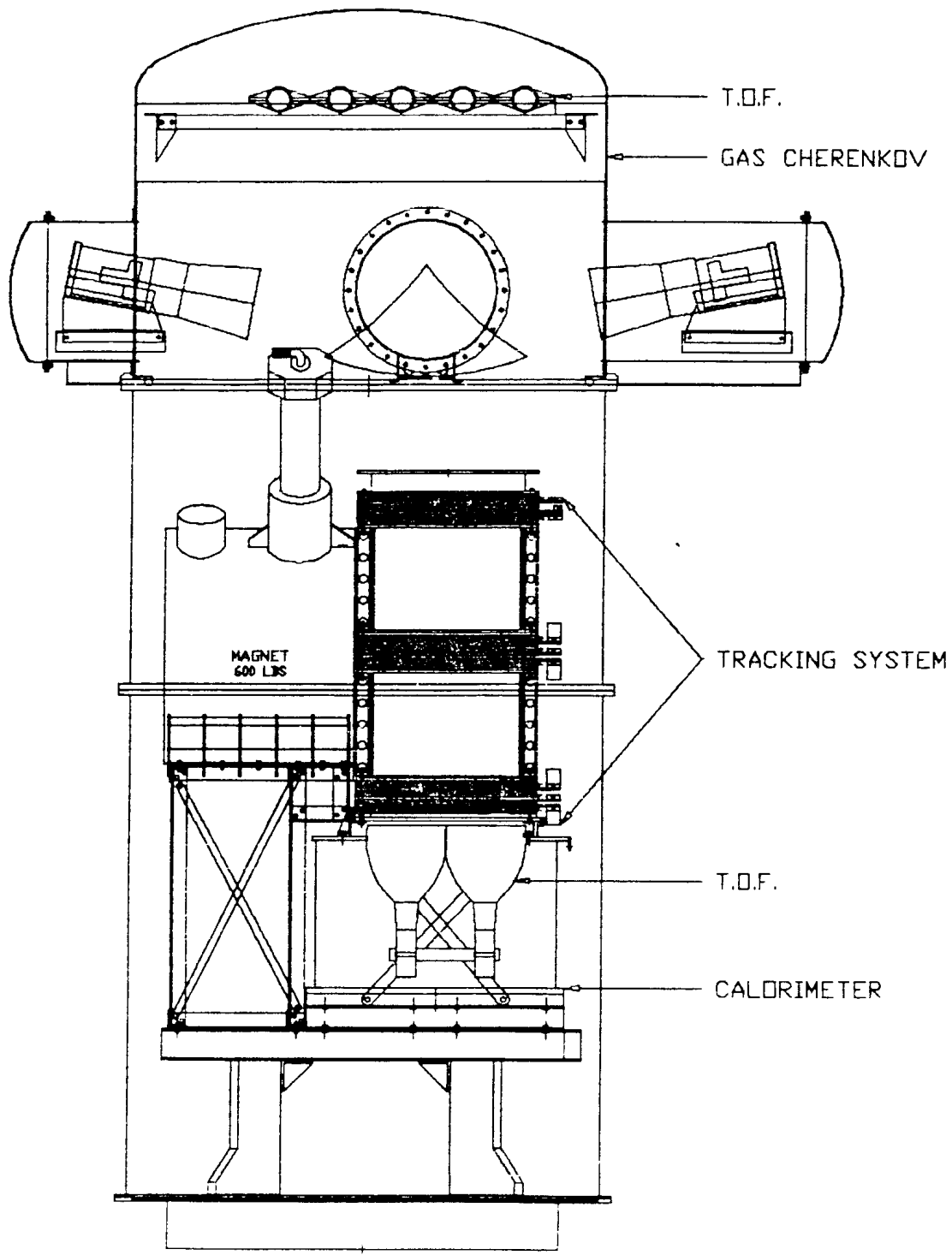,width=16cm}}
\caption{The MASS apparatus in the 1991 configuration.
\label{fapparato}}
\end{figure}

The apparatus used in the 1991 experiment was a modified version
of the MASS spectrometer flown by the same collaboration 
in 1989 
\cite{Depas}.
It consisted of a superconducting magnet spectrometer, 
a time of flight device (T.O.F.), a gas threshold Cherenkov detector and
an imaging calorimeter, 
as shown in Fig.~\ref{fapparato}.

The magnet spectrometer consisted of the NMSU single coil superconducting 
magnet and of a hybrid tracking device. The magnet, 
with 11,161 turns and a 
current of 120~A, 
gave rise 
to a field strength
of 0.1--2~T in the region of the tracking device.
The latter consisted of three groups of multiwire proportional chambers
interleaved with two drift chambers, for a total height of 110~cm. 
Each drift chamber was equipped with ten sensitive layers, each with
16 independent cells. The drift tubes
were filled with CO$_2$. The multiwire proportional chambers were 
filled with ``magic gas'', and were read by means of the cathode-coupled
delay line technique~\cite{Lacy}. 
A total number of 19 measurements
along the direction of maximum curvature and 8 measurements along the 
perpendicular direction were performed. 
The maximum detectable rigidity 
for this configuration of the 
spectrometer was estimated to be about 210 GV for singly charged 
particles \cite{Hofanti}. 

The time of flight detector consisted of two planes of scintillator
separated by a distance of 2.36~m.
The upper plane was located at the top of the apparatus. 
It consisted of two layers of scintillator, segmented into 5
paddles of 20~cm width and 
 variable length in order to match the round section 
of the payload's shell. 
The bottom plane, consisting of a single scintillator layer
segmented into two paddles, was 
located below the tracker system and above the calorimeter. 
A coincidence between the 
signals from the two planes produced the trigger for 
data acquisition. 
The signals from each paddle of scintillator were independently
digitized for time of flight measurements as well as for
pulse height analyses. \\
The Cherenkov detector consisted of a 1~m tall cylinder 
of Freon~22 at the pressure of 1~atm. 
A four-segment spherical mirror focussed the light 
onto four photomultipliers. 
The threshold 
Lorentz factor for Cherenkov 
emission was $\gamma_{th}\approx$ 25. \\
The calorimeter consisted of 40 layers, each having
64 brass streamer tubes. 
Tubes from adjacent layers were arranged perpendicular to one another.
The total depth of the calorimeter was 40 cm, equivalent to
7.3 radiation lengths and 0.7 interaction lengths for protons.

\section{Data Analysis} 
The general features of the data analysis procedures were 
the same for the three studies illustrated here.
Nevertheless, we used different sets of criteria for selecting
different particles, due to the different kinds of background
events to be eliminated and the extent of the rigidity over which the 
analysis was carried out in each of these cases.
Additional difficulties for the ascent analysis 
arise because of the possible shocks during the launch and of 
the rapidly changing environmental conditions with altitude,
namely, atmospheric pressure and temperature.
We accurately monitored the instrumental conditions 
continuously during the ascent in order to make sure that the detector 
performances did not change significantly during the data acquisition.
Further, in the case of the ascent analysis, the relative intensity 
of different particles change with altitude, and this  might
mimic instrumental drifts. 
Because of
these reasons, we used a stringent selection for ascent muons, in
such a way to make use of the full information recorded for each event.
A great deal of effort was put into checking the consistency of the
ascent selection with the muon analysis at float, which has 
been illustrated separately \cite{Codino}. \\
The proton and helium events from the float file were identified 
by selecting charge 1 and 2 particles by means of the scintillator signals. 
The selection of muon events from the ascent file were mainly 
obtained by identifying
singly charged particles which did not interact in the calorimeter.
The track reconstruction in the spectrometer allowed  
the sign of charge of the particles to be determined. 
Low-energy muons were discriminated from protons by means of 
the time of flight measurement.
Details of the event selection and analyses are 
described in the following sections.

\subsection{Event Reconstruction}
The criteria imposed for the selection of good reconstructed tracks
were based on the  experience gained with this spectrometer in this
and in other
flights \cite{Hofanti,Hofappar,Goldenappar}.
Although the spectrometer had some multiple track capabilities, 
only single track events were selected for analysis. 
The criteria used for the reconstruction of events in the 
spectrometer are summarized in Table~\ref{ttracce}. This set of 
criteria was sufficient to select clean good events for the track 
reconstruction  
from both the ascent and the float samples. 
Among the tests shown in this Table, Tests~1--6 were 
introduced in order to select only good quality reconstructed tracks. 
In addition, the required consistency between the
track extrapolation to the scintillator plane and the 
position obtained from the scintillator information 
(Test~7), the requirement
that the extrapolated track pass through the 
calorimeter (Tests~8)
and the rejection of tracks intersecting  the 
lift bar of the payload (Tests~9)
removed
multiple tracks and events  generated in interactions
in the payload. 
Finally, Test~10 on  the particle velocity as determined 
with the time of flight measurement rejected albedo events.

\subsection{Proton and Helium Selection}

The identification of protons and helium events in the 
float sample was performed 
by analyzing the pulse heights of the two independent
signals, $I_1$ and $I_2$, from the top layers of scintillator. 

\begin{figure}
\mbox{\epsfig{file=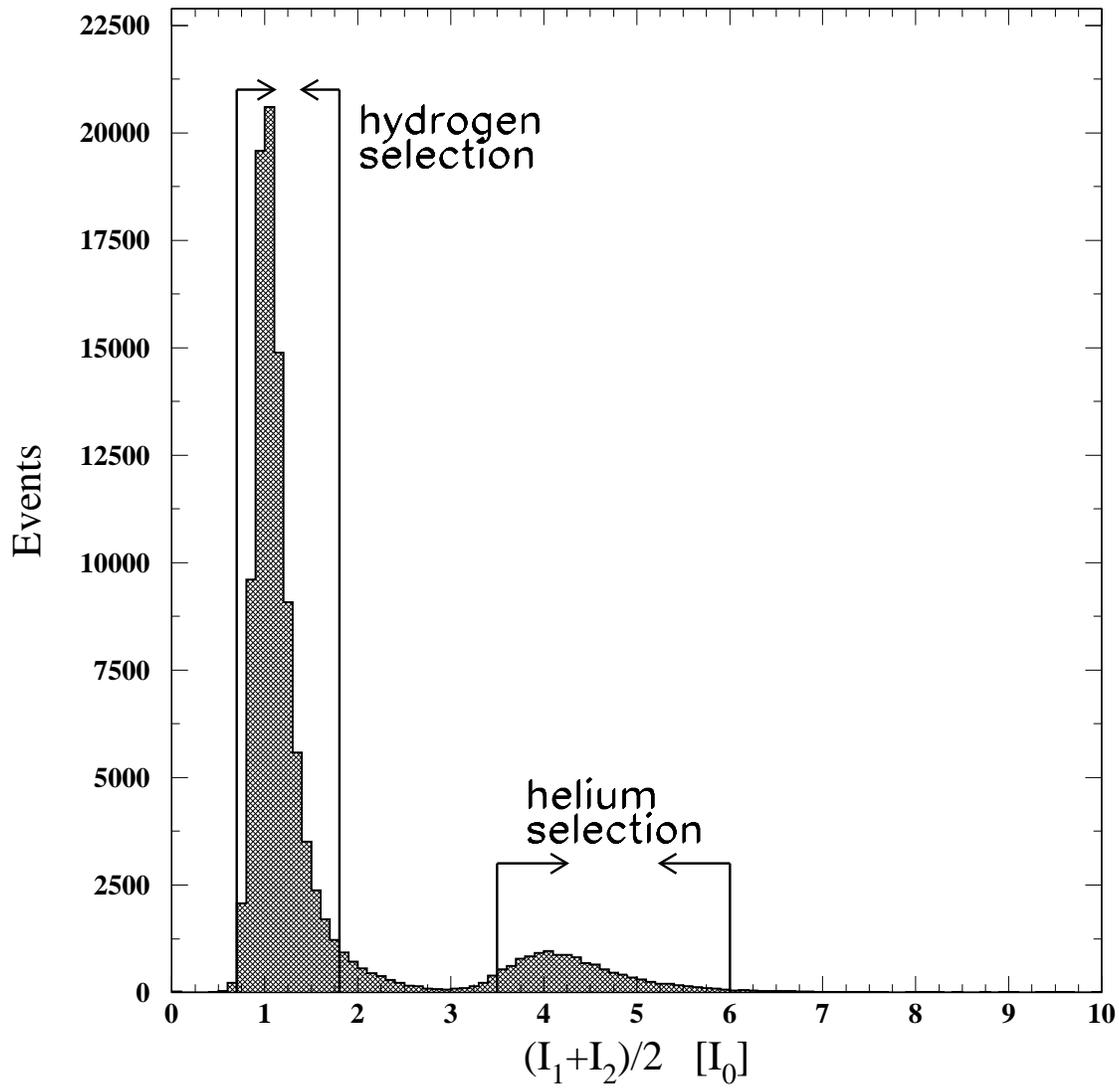,width=16cm}}
\caption{Distribution of the amplitude signals from the top 
scintillators
for high-energy positive events at float (above 3 GV). \label{f:dedx}} 
\end{figure}

The selection for charge 1 particles (protons) was:
\begin{equation}
        0.7 I_0 < \frac{I_1+I_2}{2} < 1.8 I_0 \, ,
\label{mprotscint}
\end{equation}
where $I_0$ is the mean signal from a singly charged minimum ionizing 
particle. The selection for charge 2 particles (helium) was:
\begin{equation}
        3.5 I_0 < \frac{I_1+I_2}{2} < 6 I_0 \, .
\label{mhelscint}
\end{equation}
Such  selection criteria are illustrated in Fig.~\ref{f:dedx}. 
The lower cut in the helium selection,
as given by  the above equation, 
is necessary in order to reduce 
the proton contamination in the helium sample. 
For the same reason,
consistency between the amplitudes of the two signals $I_1$ and $I_2$ 
was also required for helium selection:
\begin{equation}
         \frac{|I_1-I_2|}{\sqrt{2}} < 0.4*I_0 \, .
\end{equation}

These selection criteria are appropriate  for rigidities
above a few GV, relevant to this work.
The results concerning the proton and deuterium components 
in the atmosphere below the geomagnetic cut-off will be presented 
separately (see \cite{papini2} for a preliminary report).

\subsection{Muon Selection}

\begin{figure}
\vspace*{4.5cm}
\mbox{\epsfig{file=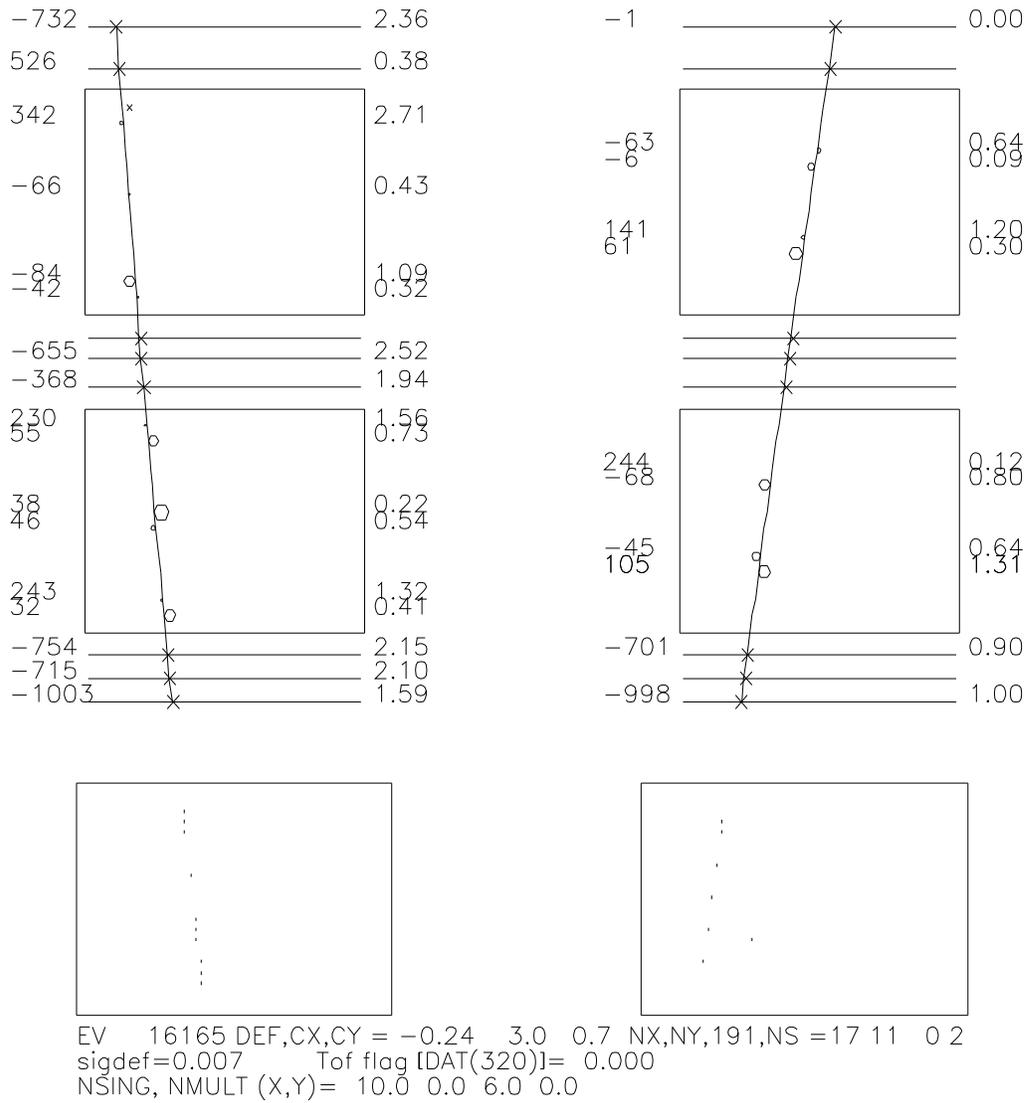,width=16cm}}
\vspace*{-4cm}
\caption{
Negative muon identified in the apparatus. 
The reconstructed event is shown along the direction of maximum bending 
in the magnetic field (left) and along the perpendicular view (right).
The estimated deflection is $\eta$=0.24 GV$^{-1}$, corresponding to 
a rigidity R=4.17 GV.
The track in the calorimeter is shown at the bottom.
\label{fevento}} 
\end{figure}

The criteria for the identification of muons of either charge 
are shown in Table~\ref{tmuoni}.
The scintillator selection (Test~1) for identifying 
singly charged
particles was the same as for the float protons 
(\ref{mprotscint}).
For the muon selection, 
the number of hits detected in the calorimeter 
was counted separately for each view, in order to account for
the different streamer tube efficiency.
Both the  minimum number of signals and the number 
of multiple hits refer to the hits contained in a cylinder  of 
radius  of 5 streamer tubes along the track extrapolation 
in the calorimeter, corresponding to about 3 Moli\`ere radii.
In particular, Test~4 is a powerful means of rejecting
electrons \cite{Grimani89}. 
An event identified as a negative muon by means of such selection 
is shown in Fig.~\ref{fevento}. \\ 
The Cherenkov signal and the time of flight information
were used  for background rejection. 
Test~5 was imposed to remove the low-energy electrons and 
positrons misidentified in the calorimeter. 
Test~6 rejects low-energy protons from the positive muon 
sample by a test of the squared mass $m^2$ 
which, once the charge $Ze$ is known, can be estimated 
from the magnetic deflection $\eta$ and the velocity $\beta$ as: 
\begin{equation}
m^2  =
\frac{\frac{1}{\beta^2}-1}{\eta^2} \times \frac{Z^2e^2}{c^2} \, . 
\label{mmassa}
\end{equation}
No time of flight test was required below 
0.65 GV, since low-energy protons are efficiently
rejected by the scintillator pulse height discrimination. \\

\subsection{ Background Estimates and Corrections}

\subsubsection{Proton and Helium Analysis}
Protons 
are the main component of primary cosmic rays. As a consequence,
the possible background from light particles, namely, positrons, 
muons and pions, is expected to be small 
above the geomagnetic cut-off and is not customarily subtracted 
from the measurement.
Therefore, no correction for such background 
events
has been performed on the proton measurements in this 
investigation.
The contamination from helium events in the selected proton 
sample is negligible. 
Further, no attempt was made to separate the isotopes of  
protons and helium events, even at low energies.
In the case of helium selection there could be
a small proton contamination due to the Landau 
fluctuations of the energy released in the scintillator layers from the 
large flux of protons.
This background was evaluated by studying a sample of protons selected 
by means of the pulse height signals in the bottom scintillator layer. 
The contamination, in the whole energy range, was found to be
 less than 2\% and for each energy bin the number of the estimated 
background protons in the helium sample was subtracted. 

\subsubsection{Muon Analysis}
The possible sources of background events, which might simulate 
moun-like events, 
are listed in Table~\ref{tcontaminazione}. 
Also shown in this table are the most efficient rejection criteria 
to eliminate the background events and
the estimated levels of residual contamination.

\begin{figure}
\mbox{\epsfig{file=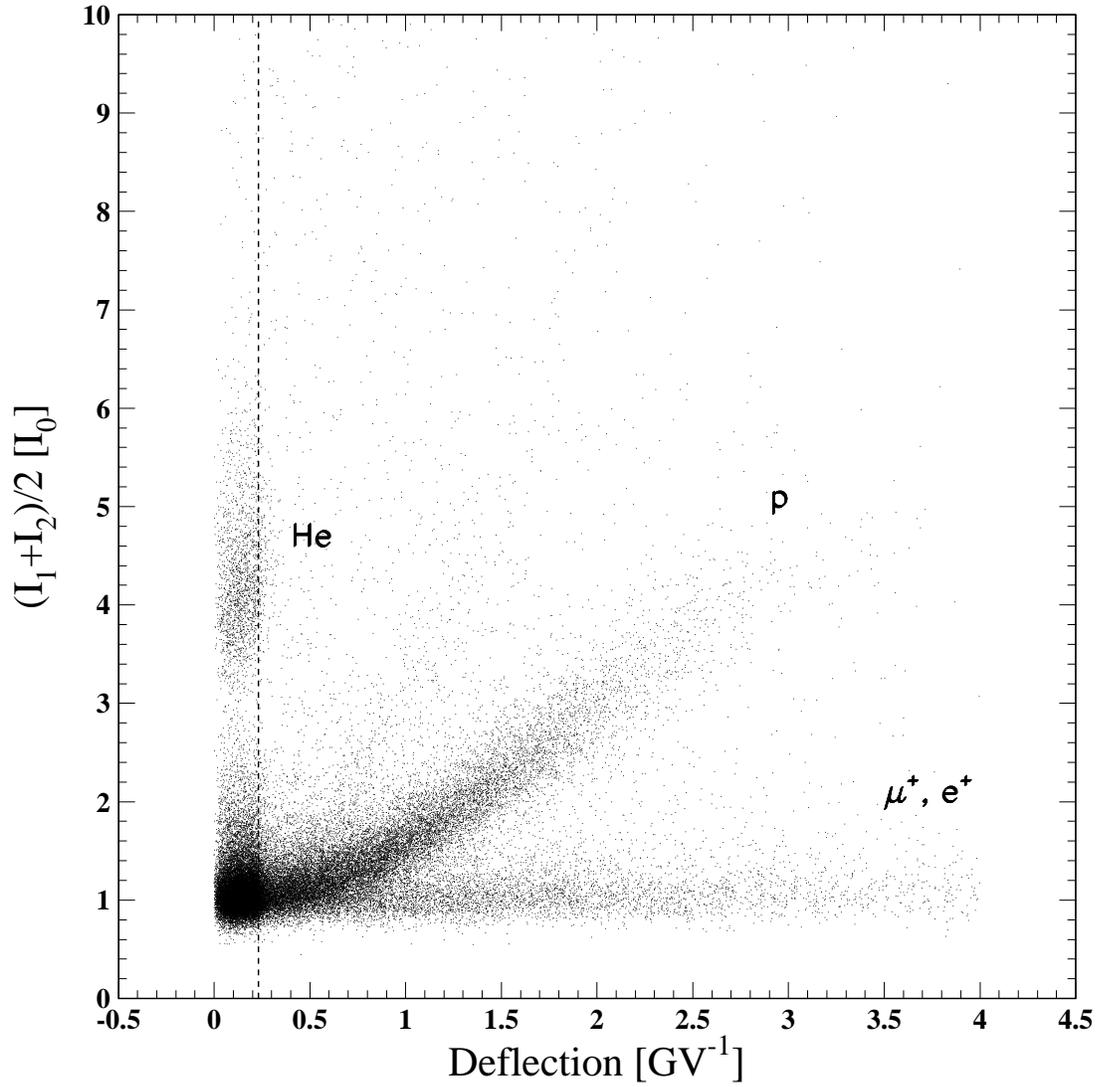,width=16cm}}
\caption{Mean pulse height in the top scintillators plotted 
as a function of 
deflection for the positive particles taken during the ascent.
The scintillator signal has been normalized to the average pulse height 
from singly charged minimum ionizing particles, $I_0$. 
A dashed line shows the value of the vertical geomagnetic cut-off
for this experiment. The effect of
the cut-off can be seen in the suppression of 
the low-energy helium component.
\label{fprimari}}
\end{figure}

Albedo events are upward-going particles which 
simulate a curvature of opposite sign in the spectrometer. 
They are either produced as large angle secondaries in interactions by 
hadrons incident at large zenith angles or by hard scatterings.
However,
we found only 9 upward-going events in the whole ascent sample. 
They were easily removed by means of the time of flight measurement
(Test~10 of Table~\ref{ttracce}). \\
The degree of possible electron contamination varies with
altitude and energy because of the different development of 
the electron and muon fluxes in the atmosphere. 
For energies $\lesssim$ 1 GeV, 
the worst conditions for the relative ratio of muon to electron 
flux is expected at less than 100 g/cm$^2$, 
where the muon flux is still increasing with atmospheric depth and
the electron flux has already reached its maximum \cite{Daniel}.
Low-energy electrons and positrons misidentified as muons
in the calorimeter were rejected from the muon sample 
by means of the Cherenkov selection shown
in Test~5 of Table~\ref{tmuoni}.
We used the number of Cherenkov-identified electron events to estimate 
the upper limit to the altitude-dependent residual contamination 
as given
in Table~\ref{tcontaminazione}. \\
Spillover events are particles whose charge sign is misinterpreted 
in the magnet spectrometer.
This source of background needs to be considered
 for the negative 
muon sample because of the large number of protons at high 
altitudes. 
As a consequence of the high performances of the magnet 
spectrometer, spillover is expected to be only a negligible source of 
background in the momentum range of this investigation. 
In fact, we carried out a simulation \cite{tesipap}, which takes 
into account the details of the magnetic field in the spectrometer 
and detector response. We found that the spillover background can not 
be more than 1\% of the negative muon events even near the float altitude 
and at the highest rigidity bin, where one might expect some contribution.
\\  
Background due to pions and kaons is the major
concern for the muon measurements, because it is not possible 
to identify mesons that do not interact in the calorimeter.
From
theoretical expectations for the pion and kaon
fluxes in the atmosphere \cite{Stephenspion,Badhwar}, we estimated that 
for muon momenta less than 10 GeV/$c$
pions do not contaminate significantly the muon measurements 
at depths larger than 200 g/cm$^2$, while an
altitude-dependent pion contamination of the order of 1-2\% can 
not be excluded at smaller depths. The fraction of contaminating pions may 
be larger at larger particle momenta. 
The kaon contamination is  negligible everywhere. 
There is also the possibility that locally produced particles, namely 
secondaries produced by hadrons interacting
in the shell or in the lift bar above the 
payload, may be detected as single muon-like events.
In order to reject such events, 
we excluded from the analysis all the tracks whose extrapolation
did intersect the lift bar. In addition, we placed severe requirements 
on the reconstructed tracks, as illustrated previously, by 
which multiple particles from an interaction that are incident within 
the instrument can be rejected. 
From an analysis of simulated events, 
we estimated that the possible residual 
contamination 
from locally produced particles is negligible, except at very low 
enrgies and at small atmospheric depths.
In order to evaluate  
the possible extent of contamination in this region,
we checked the number of negative events which 
were selected as muons in the rest of the apparatus and passed
a pion selection criterion in the calorimeter. From this fraction
and from the estimated efficiency for such a test to 
detect pions, we estimated that a contamination by locally 
produced particles at an extent up to 20\% can not be excluded for
muons below 1 GeV/$c$ at small atmospheric depths.
The fraction of such events
decreases rapidly with increasing atmospheric depth
and we found that 
it may not exceed 5\% at depths larger than 50 g/cm$^2$.
It should be emphasized that this procedure can allow us only to set
the upper limit to the contamination due to this source of background.
Therefore no further correction was made to the data. 
\\
Finally, the proton background is important for the positive 
muon measurements, since their flux rapidly increases 
with increasing altitude. 
Primary protons exponentially attenuate in the atmosphere with an 
absorption length $\Lambda\sim$ 120 g/cm$^2$ \cite{Gaisserbook}. 
This occurrence places a serious constraint on the range of
atmospheric depth over which positive muon measurements are possible.
However, the situation is different at low energy because of the 
geomagnetic suppression of primaries, as can be seen from the 
helium spectrum shown   
in Fig.~\ref{fprimari}. 
The low-energy proton component  therefore has to be of a secondary 
nature. This can also be seen in
the altitude distribution of such events 
\cite{Tesi}. The geomagnetic suppression allows us to perform a 
low-energy proton rejection by means of the 
squared mass tests
listed in Table~\ref{tmuoni}.

\subsection{Geometric Factor and Efficiencies}
\subsubsection{Geometric Factor and Global Efficiencies}
\begin{figure}
\mbox{\epsfig{file=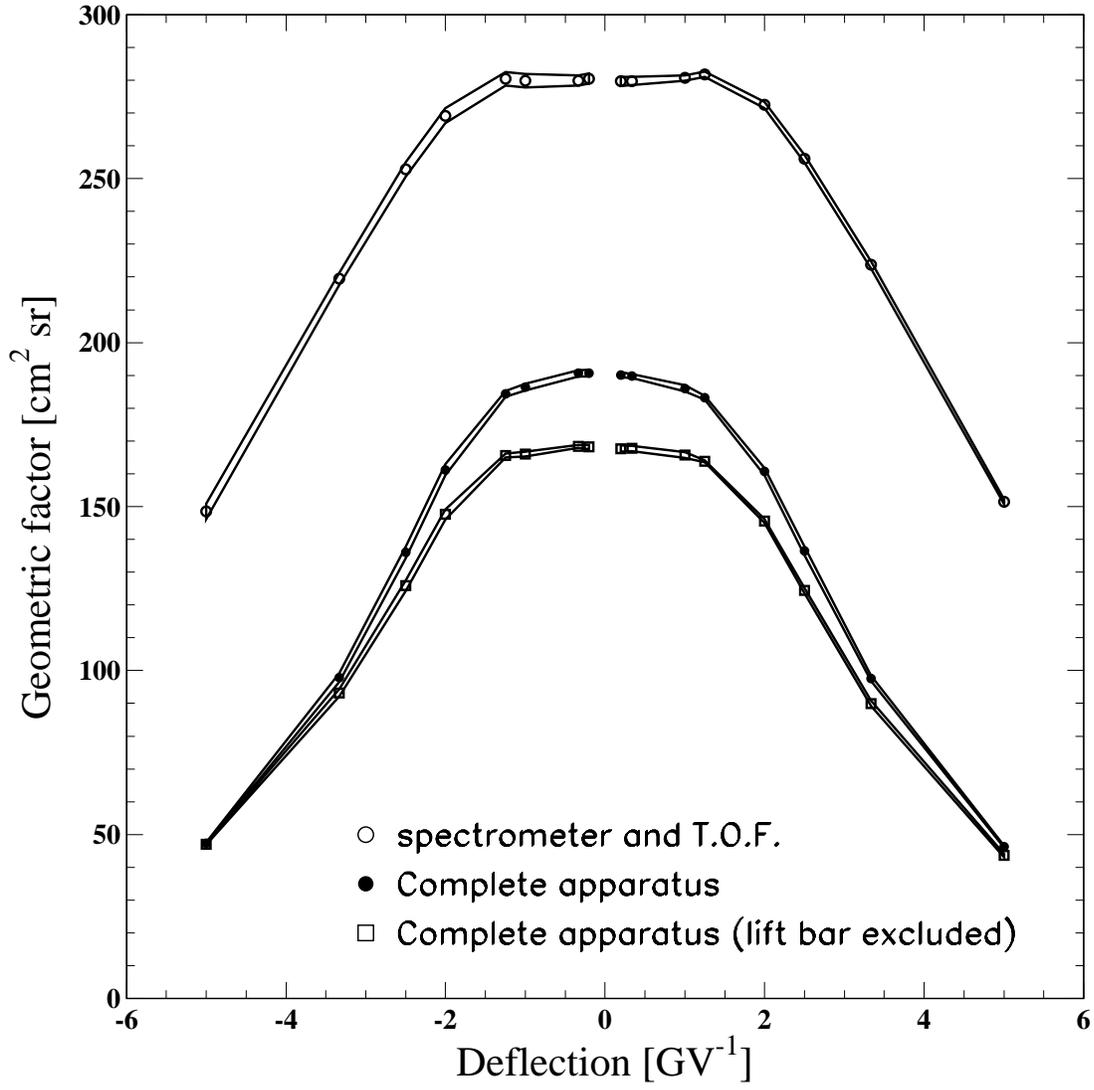,width=16cm}}
\caption{Geometric factor for different acceptance criteria.
The estimated uncertainties are also shown. 
\label{fgeometric}}
\end{figure}

\begin{figure}
\mbox{\epsfig{file=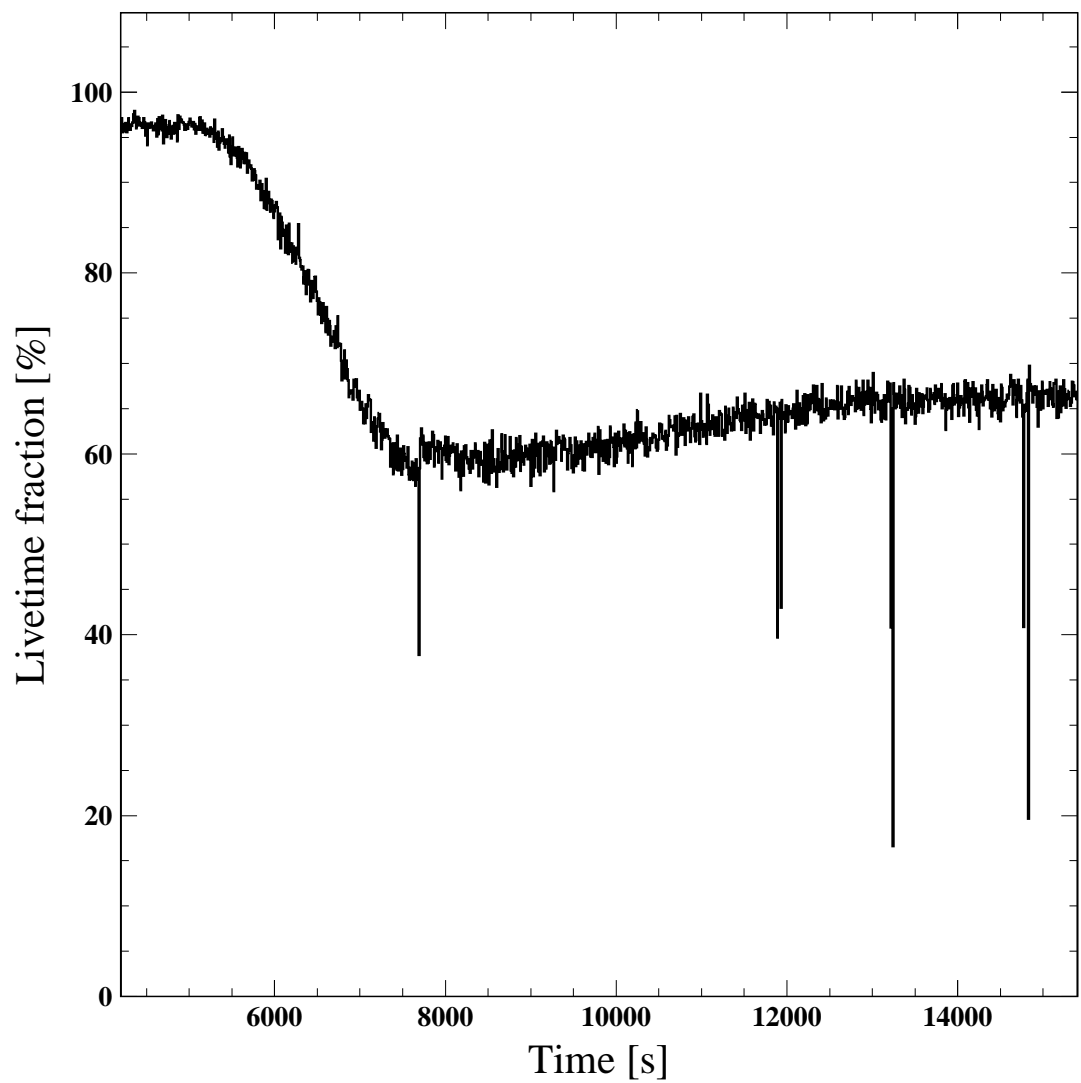,width=16cm}}
\caption{Livetime fraction during the ascent of the flight.
\label{flivetime}}
\end{figure}

\begin{figure}
\mbox{\epsfig{file=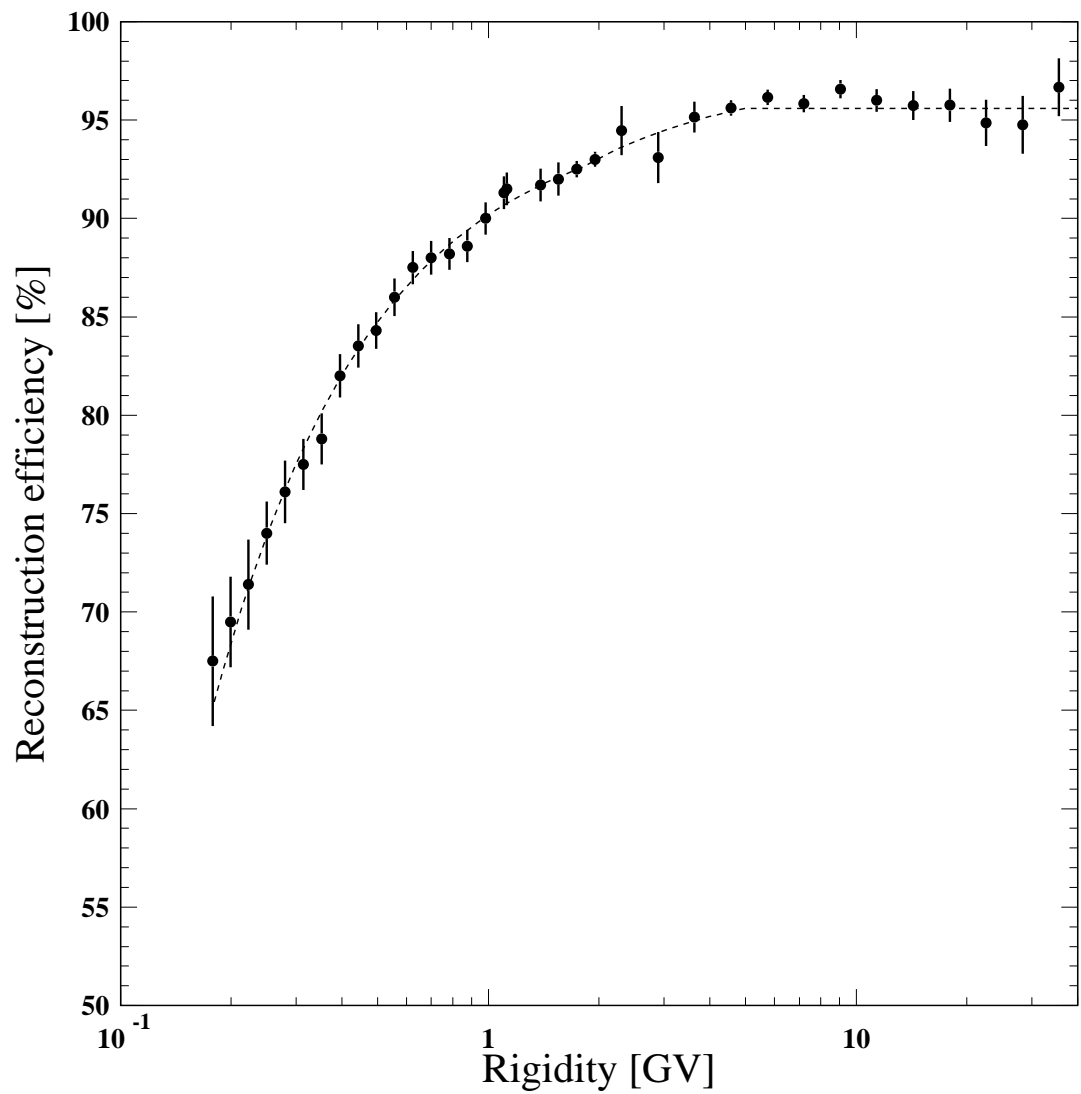,width=16cm}}
\caption{Spectrometer track reconstruction efficiency for muon particles. 
The dashed line shows 
the best-interpolation curve used in the flux
calculations.
\label{fspettrometro}}
\end{figure}

The geometric factor of the apparatus was estimated by means
of two independent codes for the containment conditions listed in
Table~\ref{ttracce}. The accuracy of such calculations was estimated 
to be better than 1\%. \\
Particles generated at the top of the apparatus were followed 
down to the bottom of the calorimeter and then traced  up to 
the level of the lift bar. 
By requiring that the track should not intersect
the suspension bar,
about 10\% of the events above 1~GV that cross the whole detector
were eliminated.
The deflection dependence of the geometric factor 
for the different cases is shown in Fig.~\ref{fgeometric}.
The small difference between positive and negative particles at high 
deflection is due to a mechanical asymmetry of the magnet with 
respect to the detector stack.\\
The following global efficiencies were introduced in each analysis:
(a) a trigger efficiency of 0.825$\pm$0.010 measured in a 
ground test before the launch; 
(b) a time-dependent livetime fraction, which varied during the 
ascent as shown in Fig.~\ref{flivetime} and reached a value of 
0.66$\pm$0.01 at float;
(c) a rigidity dependent reconstruction efficiency, shown in
Fig.~\ref{fspettrometro} for muons. While the reconstruction
 efficiency, at high energy, is the same for protons and muons, 
it is significantly lower for helium nuclei. Above the geomagnetic cut-off 
the reconstruction efficiencies were nearly 
constant; they were 0.959$\pm$0.012 for protons and muons 
and 0.917$\pm$0.032 for helium. No dependence was found 
on the sign of charge for muons.

\subsubsection{Proton and Helium Selection Efficiencies}
The scintillator efficiencies for 
charge 1 (protons and muons) and charge 2 (helium)
particle selection
were determined using samples of events
tagged by the bottom scintillator detector;  this
information was
not used in the analysis for the event selection.
This technique allows a reliable 
evaluation of the selection efficiencies. 
We estimated a selection efficiency of 0.945$\pm$0.001 for
protons and muons and 0.882$\pm$0.022 for helium nuclei. 

\subsubsection{Muon Selection Efficiencies}
In addition to the above efficiencies, 
the following selection efficiencies were considered in 
estimating the muon fluxes: 
(a) a calorimeter efficiency of 0.888 $\pm$ 0.008;
(b) the requirement of the presence of the calorimeter information 
introduced a further efficiency of 0.851 $\pm$ 0.005, because of 
some acquisition failures;
(c) the Cherenkov test at low energy was passed by muons 
with an efficiency of 
0.998 $\pm$ 0.001;
(d) the efficiency for the time of flight selection of low-energy 
positive muons was found to be 0.992 $\pm$ 0.004 and 0.908 $\pm$ 0.014, 
respectively in the 0.65--1.25 GV and 1.25--1.5 GV rigidity ranges. \\
The overall efficiency for muon selection was therefore 
a function of time and energy.
It ranged for negative muons  from a minimum of 0.298 $\pm$ 0.006
at 0.3 GeV/$c$ at maximum deadtime 
to the value of 0.539 $\pm$ 0.011 above 4~GeV/$c$ and
at minimum deadtime. 
The detection efficiency for positive muons was slightly lower 
because of the additional squared mass selection criterion.

\section{Results}
\subsection{Muon Results}
\begin{figure}
\mbox{\epsfig{file=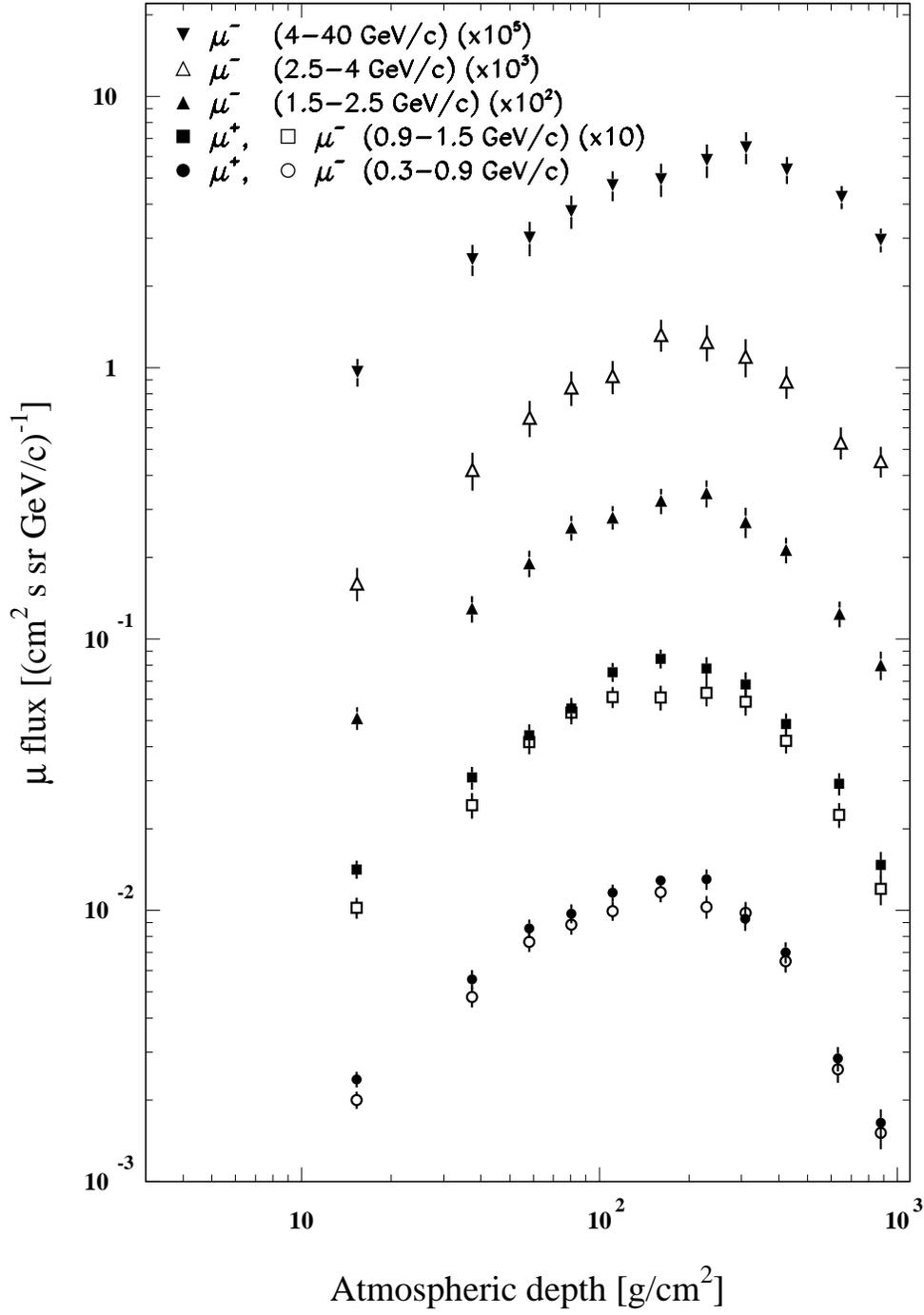,width=14cm}}
\caption{Flux growth curves for negative muons 
in the 0.3--40 GeV/$c$ momentum range. 
Positive muon results are shown in the 0.3--1.5 GeV/$c$ momentum interval.
Some of the distributions have been scaled as indicated. 
\label{fcurve}}
\end{figure}

\begin{figure}
\mbox{\epsfig{file=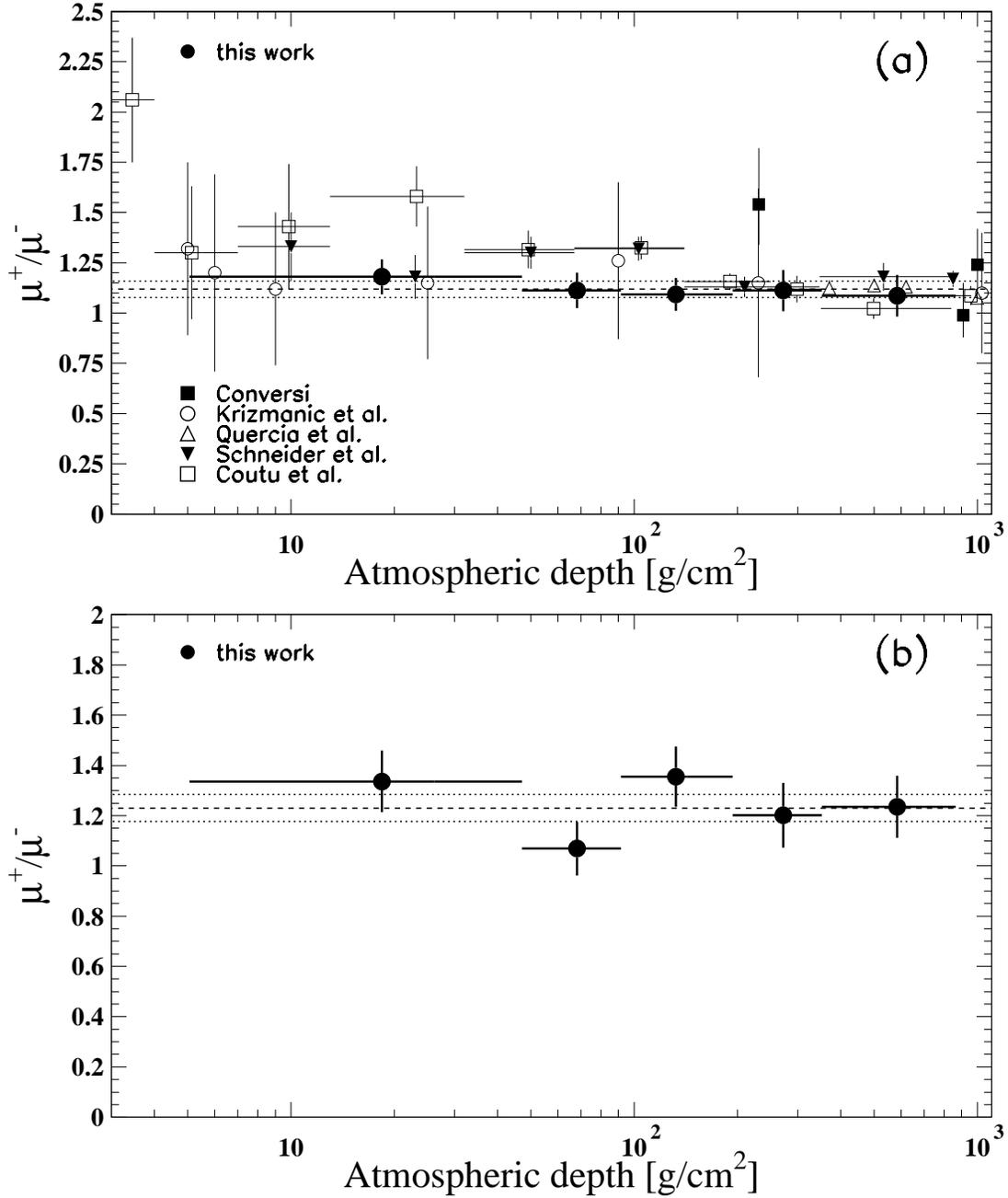,width=16cm}}
\caption{Muon charge ratio in the (a) 0.3--0.9 GeV/$c$ and 
(b) 0.9--1.5 GeV/$c$ momentum intervals with changing atmospheric depth. 
The dashed lines show the weighted average values from this 
experiment, and the dotted lines the corresponding 1$\sigma$ intervals.
Results from previous experiments are also shown: 
Conversi (0.315--0.348 GeV/$c$)~\protect\cite{Conversi},
Krizmanic {\it et al.} (0.42--0.47 GeV/$c$)~\protect\cite{IMAX},
Quercia {\it et al.} ($\ge$460 MeV)~\protect\cite{quercia},
Schneider {\it et al.} and Coutu
{\it et al.} (0.3--0.9 GeV/$c$)~\protect\cite{HEAT95,HEAT97}.
\label{frapporto}}
\end{figure}

\begin{figure}
\mbox{\epsfig{file=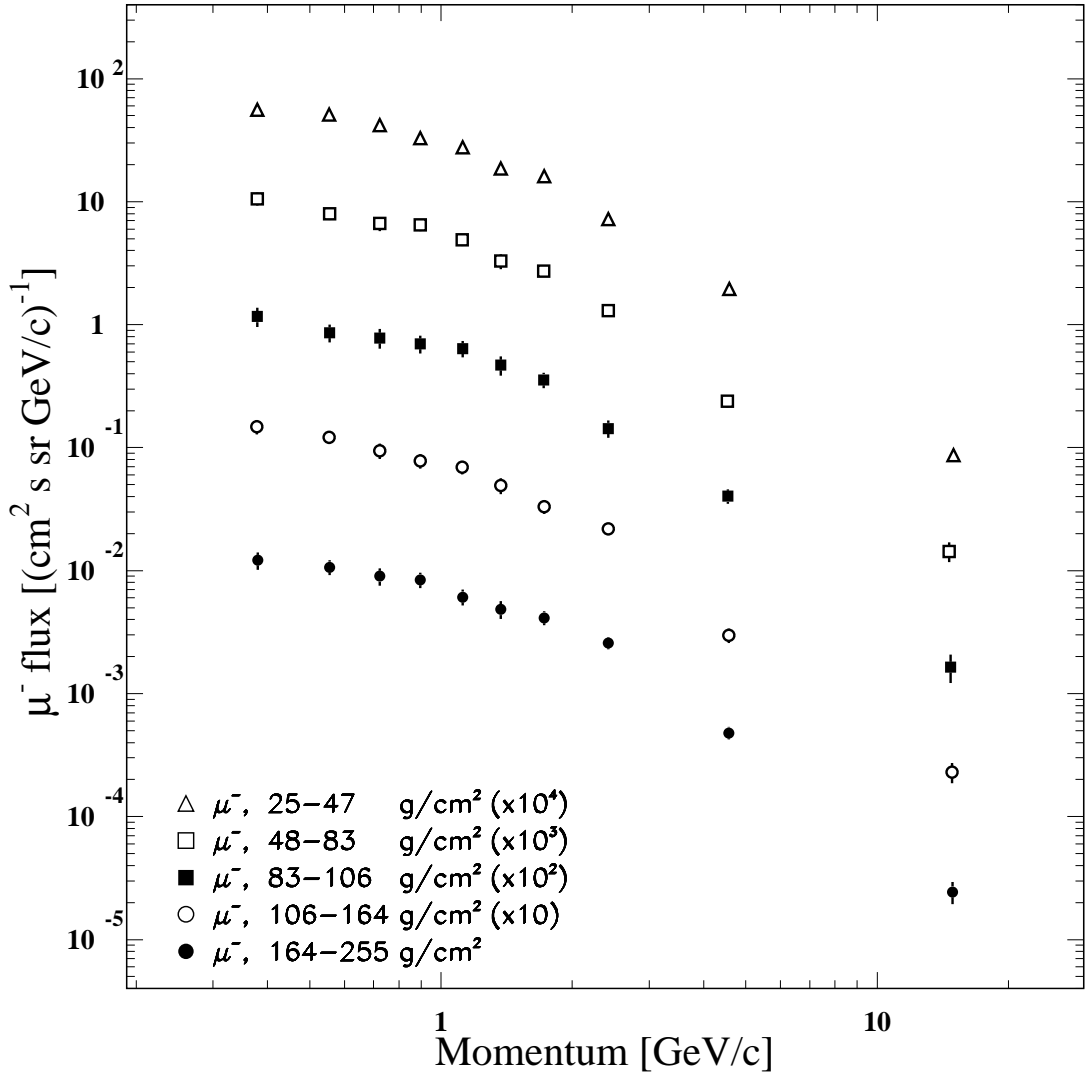,width=16cm}}
\caption{Negative muon momentum spectra in different depth intervals.
Some of the distributions have been scaled as indicated. 
\label{fspettri}}
\end{figure}

\begin{figure}
\mbox{\epsfig{file=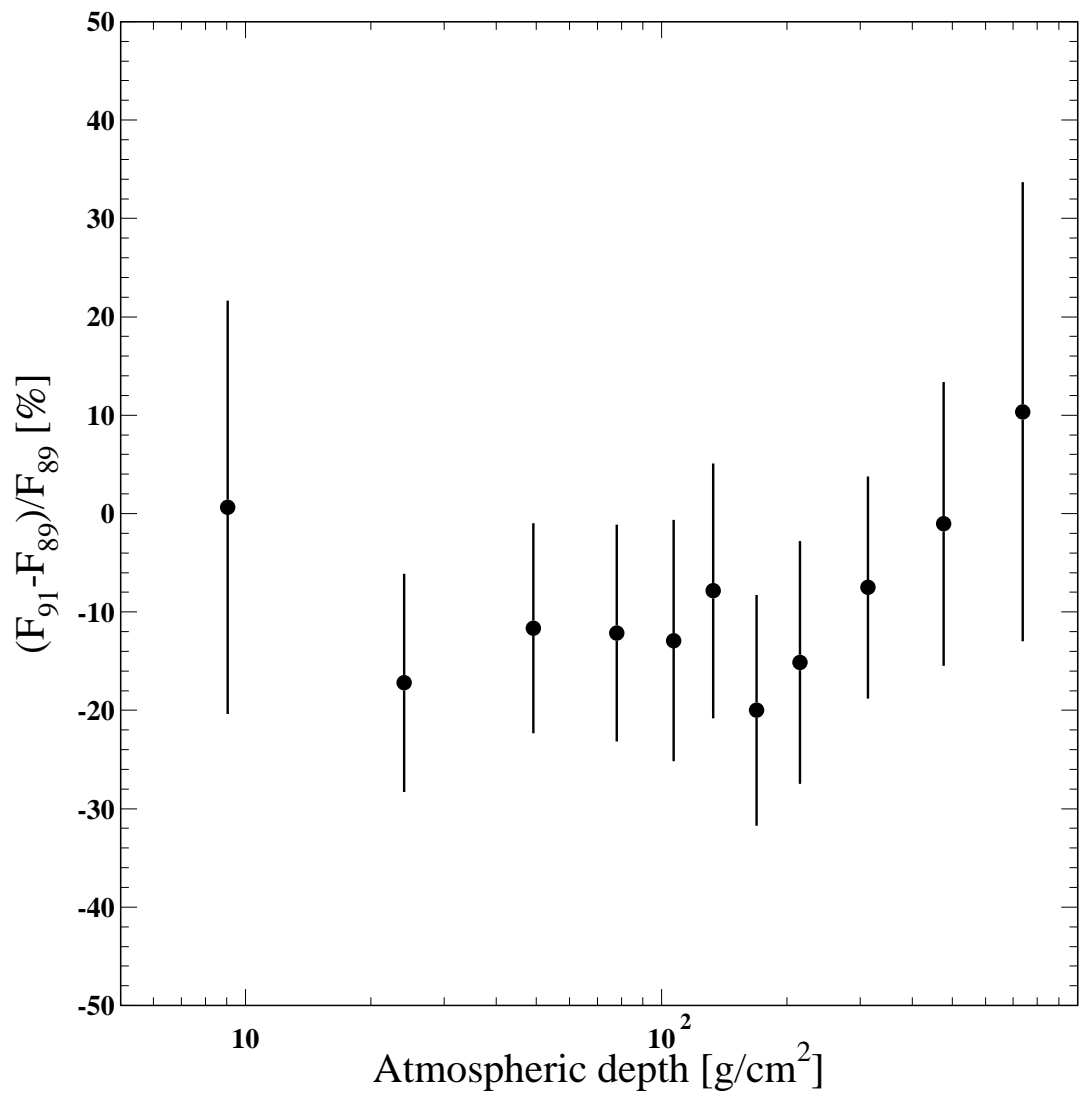,width=16cm}}
\caption{Differences in the 0.3--1 GeV/$c$ negative muon 
flux measured in this experiment with respect to the 1989 experiment
\protect\cite{PRD}. 
\label{fdiff}}
\end{figure}

With the selection
described
in the previous sections, we 
selected a sample of 4,471 negative and 
2,856 positive muons
distributed in the atmospheric depth range of 5-886 g/cm$^2$.
As previously mentioned, the momentum range investigated for negative 
muons was from 0.3 to 40 GeV/$c$, while positive muons were selected 
in the 0.3--1.5 GeV/$c$ interval. \\
We followed the same procedure developed for our previous 
analysis~\cite{PRD} for the reconstruction of the flux growth curves 
in the atmosphere. 
In particular, a parametrization of the form:
\begin{equation}
\Phi(X) = k X e^{-X/\Lambda} 
\label{mprofilo}
\end{equation}
was adopted in order to describe the dependence of the muon flux
in the different momentum intervals upon the 
atmospheric depth $X$, where
$k$ and $\Lambda$ are varied
to fit the data. 
Results on the depth dependence of muons of either charge are shown in
Fig.~\ref{fcurve}, and are also given
in Table~\ref{tcurve}: it may be noted that the positive and negative 
curve shapes do not show any noticeable
difference. \\
Fig.~\ref{frapporto} shows the muon charge ratio 
in the atmosphere in two different energy intervals. It can be 
noticed that our results do not show any definite trend of the charge ratio 
changing with atmospheric depth.
On the other hand, it may be pointed out that the
depth-averaged value of the $\mu^+/\mu^-$ ratio
increases with increasing momentum of the particles,
being 1.12 $\pm$ 0.04 and 1.23 $\pm$ 0.05 respectively in the
0.3--0.9 and 0.9--1.5 GeV/$c$ momentum bins.
These values are consistent with the ratio measured 
at float in the same experiment~\cite{Codino}. 
Fig.~\ref{frapporto} also shows that, while there is a general agreement 
among results in the low energy bin at large atmospheric depths, there is 
noticeable difference at low altitudes below 100 g/cm$^2$. In addition to 
the results shown in Fig.~\ref{frapporto}, results are also available 
at very small atmospheric depths. The CAPRICE experiment 
reported an average value of 1.64$\pm$0.08 
between 0.2 and 2.3~GeV/$c$ at 3.9~g/cm$^2$
of residual atmosphere~\cite{CAP94}, while a ratio of 1.26$\pm$0.12 
for 0.3--1.3 GeV/$c$ muons was previously found at 11 g/cm$^2$~\cite{bogo}.
It is not clear from the 
literature how much of the differences  
in the observed 
ratio could be ascribed
to the different experimental conditions.
\\
The measured spectra of negative muons 
at different depths  between 25 and 255 g/cm$^2$ 
are shown in Fig.~\ref{fspettri} and also
in Table~\ref{tspettri}:
The results in Table~\ref{tcurve}
show that, in spite of the differences in the growth pattern of 
the muon flux for different momentum intervals, the estimated value of the 
effective atmospheric depth (FAD) do not differ by more than 1\% at all 
depths, except at the largest depth interval. 
Above 1.5 GeV/$c$, the negative muon spectra 
may be parametrized as power-laws 
with a power index of 2.45$\pm$0.05, 
almost independent of the atmospheric depth, 
and in a close agreement with our previous observations in~\cite{PRD}.
A comparison between these two measurements
shows that the normalizations of the two sets of
results 
are in a good agreement
in the 1--8 GeV/$c$ interval.
A comparison between these two experiments at lower energy is less 
straightforward, due to the different conditions of solar modulation and 
geomagnetic cut-off of the two experiments. 
As shown in Fig.~\ref{fdiff}, we measured a significant deficit 
of low-energy muons in the 1991 flight with respect to the 
1989 experiment
over a large range of atmospheric depth.

\subsection{Proton and Helium Results}
From the events recorded at the float, we have selected 
118,637 proton events and 15,207 helium events for the analysis.
These events were collected over a period of 35,330 s. After 
subtracting the estimated background, the number of events were 
corrected for the selection efficiencies.   
The flux for each selected energy bin at the spectrometer level 
was estimated using the time of observation
and the calculated geometric factor. 
In the case of protons, we 
have chosen the rigidity range between 3.3 and 100 GV, where the 
contribution from the atmospheric secondaries is small. The helium 
spectrum was investigated between 3 and 100 GV.

The estimated flux values at the spectrometer level
were corrected to the top of the payload  by taking into account 
inelastic interactions and ionization energy loss in the detectors
above the spectrometer (namely, the plastic scintillator counters and
the gas Cherenkov detector) and in the aluminum dome of the payload. 
The proton flux was then extrapolated to the top of the 
atmosphere by making use of the procedure described by  Papini {\it et al.} 
\cite{Papiniprot}, which includes the ionization and interaction losses 
as well as 
the secondary production in the residual atmosphere above the apparatus. 
In the case of the helium nuclei, in 
addition to the ionization and interaction losses, 
the production by heavy nucleus spallation was taken into account
by considering the appropriate helium attenuation length instead 
of the helium interaction length.

\begin{figure}
\mbox{\epsfig{file=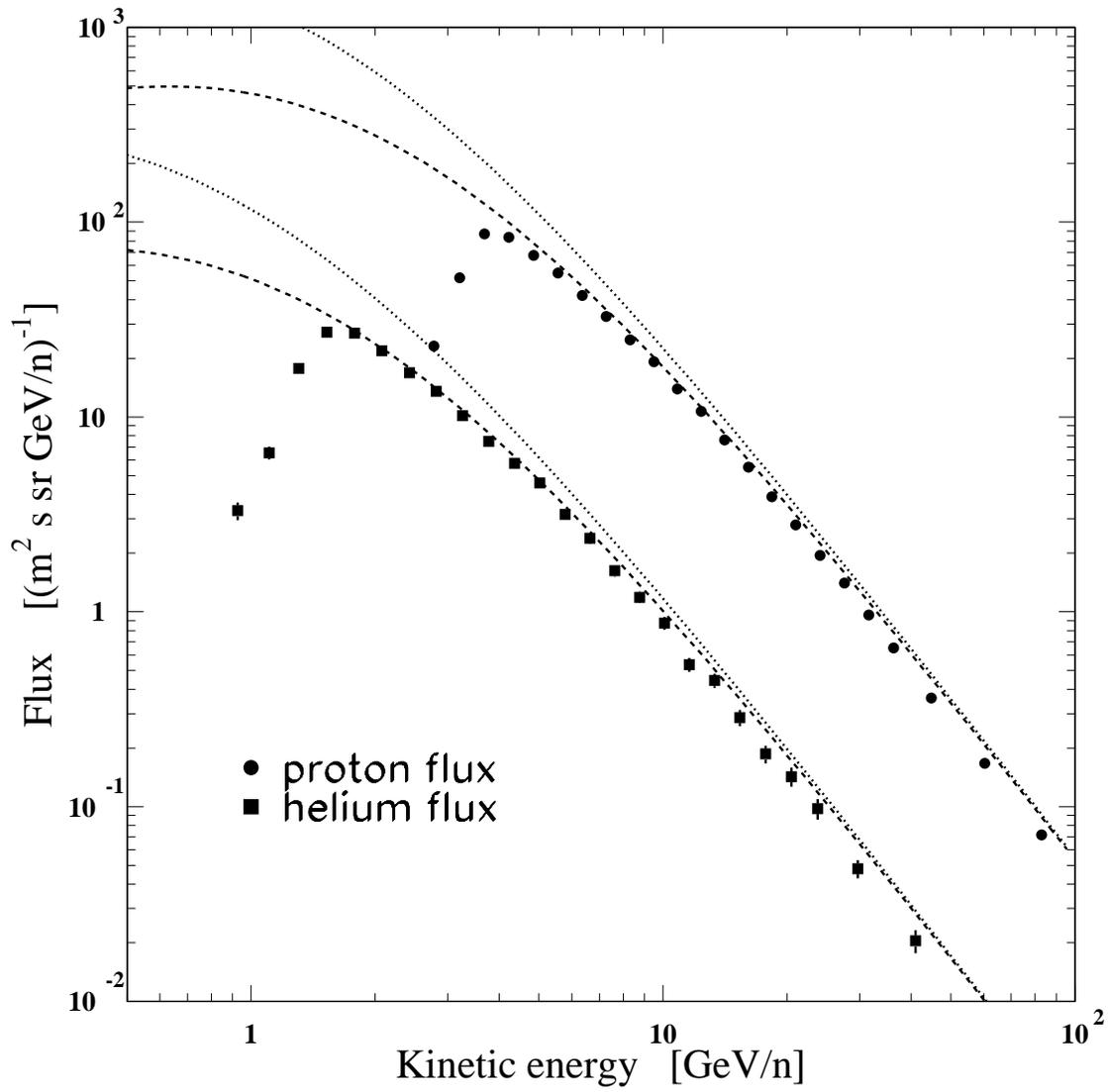,width=16cm}}
\caption{Proton and helium flux values at the top of 
the atmosphere are plotted as a function of energy.
The superimposed curves 
represent fits of a previous compilation of data 
\protect\cite{Papiniprot} for minimum (dotted) 
and maximum (dashed) of solar modulation. 
\label{f:mass2_prhe}} 
\end{figure}

The proton and helium fluxes at the top of the atmosphere are given 
in Tables \ref{t:pr_flux} and \ref{t:he_flux}. The spectral index 
$\gamma$ 
is $2.708 \pm 0.037$ for protons  above 30 GeV and 
$2.65 \pm 0.19$ for the helium flux above 15 GeV/n. 
The measured spectra are shown in Fig.~\ref{f:mass2_prhe}, where
the  geomagnetic effect is evident below 3.5 GeV for protons and
below 1.5 GeV/n for helium. 
The spectral shapes of the data, above the
geomagnetic cut-off, show that the solar modulation
effect is noticeable despite the high
value of geomagnetic cut-off for this experiment.

Because of the penumbral bands associated with the geomagnetic 
cut-off rigidities at mid-latitudes, primary cosmic rays are partially 
transmitted through the earth's magnetic field near the cut-off. In the 
following, we attempt to determine the geomagnetic transmission function, 
which is defined as the fraction  of cosmic rays of given energy to 
reach the Earth after the interaction with the geomagnetic field, from our 
observation. For this purpose, we make use of the observed helium spectrum 
rather than the proton spectrum, because at low energies the secondary 
production of protons in the atmosphere influences the measured proton 
spectrum.

\begin{figure}
\mbox{\epsfig{file=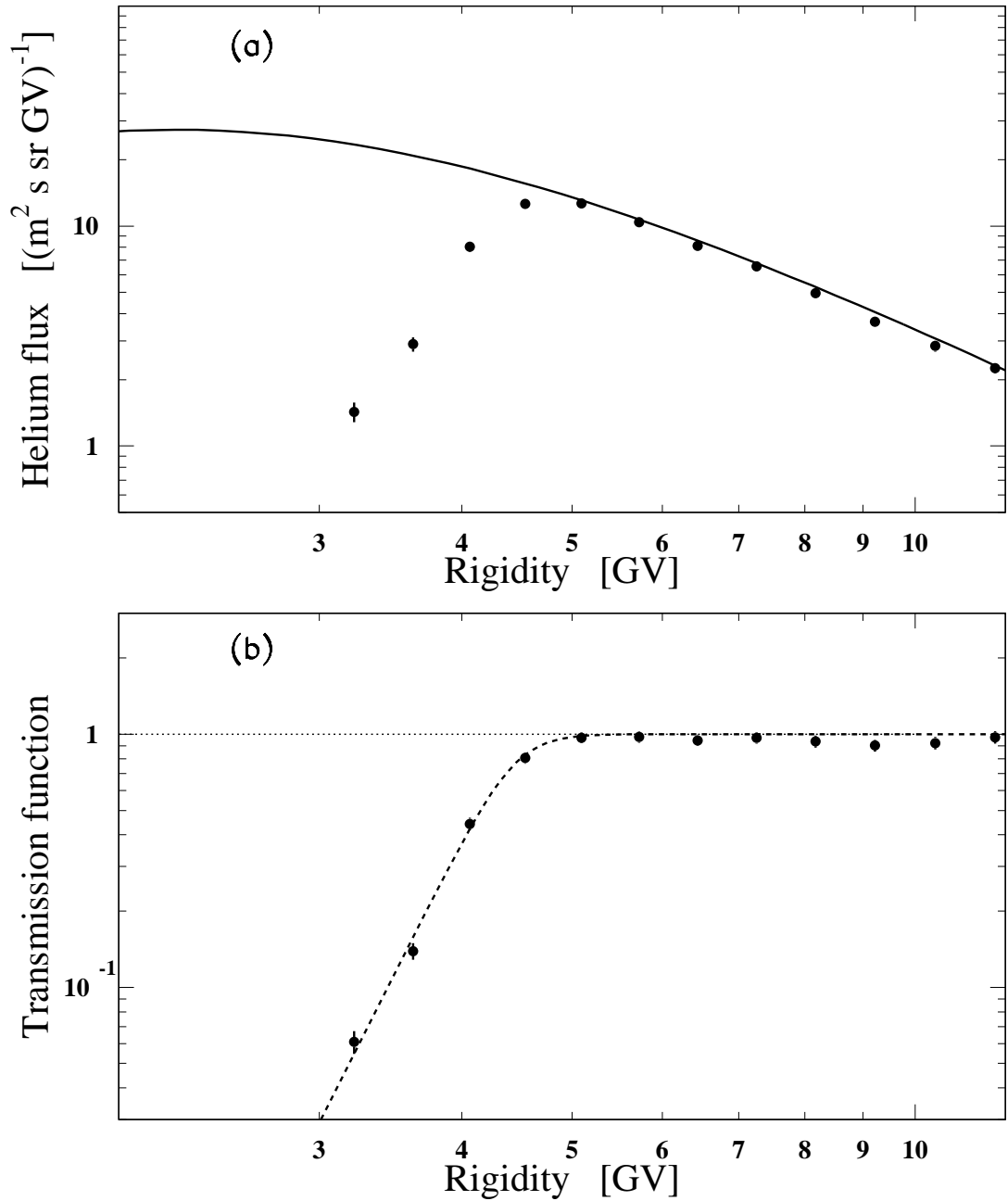,width=16cm}}
\caption{Transmission function in the geomagnetic 
field. Panel (a) shows
the  helium flux measured in this experiment
as a function of rigidity; Panel (b) shows the fitted 
transmission function as resulting from the
ratio of the experimental points and
the normalized curve given in 
Fig.~\protect\ref{f:mass2_prhe} for maximum solar
modulation. \label{f:transm}}  
\end{figure}

In Fig.~\ref{f:transm}(a) the helium flux is
shown as a function of rigidity together with the curve of 
Fig.~\ref{f:mass2_prhe} corresponding to 
the maximum of solar modulation \cite{Papiniprot}. The ratio
between the experimental points and the curve
is shown in Fig.~\ref{f:transm}(b). This ratio can be taken as
representative of the transmission function. The dashed  curve is the  
best-fit parametrization of the data with a simple curve:
\begin{equation}
GF(R)=\left\{ \left[ \left( 0.920\pm0.010 \right) \times \left( R/R_c \right) 
\right] ^{\left(-23.2\pm2.0\right)}+1 
\right\}^{\left(-0.385\pm0.040\right)} \, ,
\end{equation}
where $R_c$=4.1 GV
represents the average value of  the effective vertical 
cut-off rigidity over the flight trajectory. 
This average value has been estimated using 
the vertical cut-off map by Shea and Smart \cite{shea}. 
The position of the payload changed between 
$34^{\circ} 43^{\prime}$ and 
$35^{\circ} 29^{\prime}$ of N-latitude  
and between $103^{\circ} 38^{\prime}$ and
$104^{\circ} 25^{\prime}$ of W-longitude 
during the flight, 
with a small variation in the value of vertical cut-off. \\
It can be useful to have an analytical representation of the measured 
primary fluxes . For this purpose, it has been found 
that a simple function of the form
\begin{equation}
J(E)= a \left( E + b \mbox{e}^{-cE} \right)^{-\gamma} \times GF(E)
\label{mprimary}
\end{equation}
 can  fit the data both for proton and helium spectra. 
In (\ref{mprimary})
$a$, $b$, $c$ are 
free parameters, $\gamma$ is the slope of the spectrum at high energy, 
$GF(E)$ is the geomagnetic transmission function and $E$ is the kinetic 
energy per nucleon. 
The parameter values obtained for protons are: 
$a=11169\pm121$, $b=2.682\pm0.046$, $c=0.0950\pm0.0059$ with a 
reduced $\chi^2=1.12$;
the corresponding values for helium are 
$a=406\pm14$, $b=1.416\pm0.068$, 
$c=0.203\pm0.039$ with a reduced $\chi^2=0.51$.
We found that a parametrization (\ref{mprimary})
can represent, with the same accuracy and
in the same
energy range explored in this work, 
the observed spectra of all recent measurements
by using different values for the constants.

\begin{figure}
\mbox{\epsfig{file=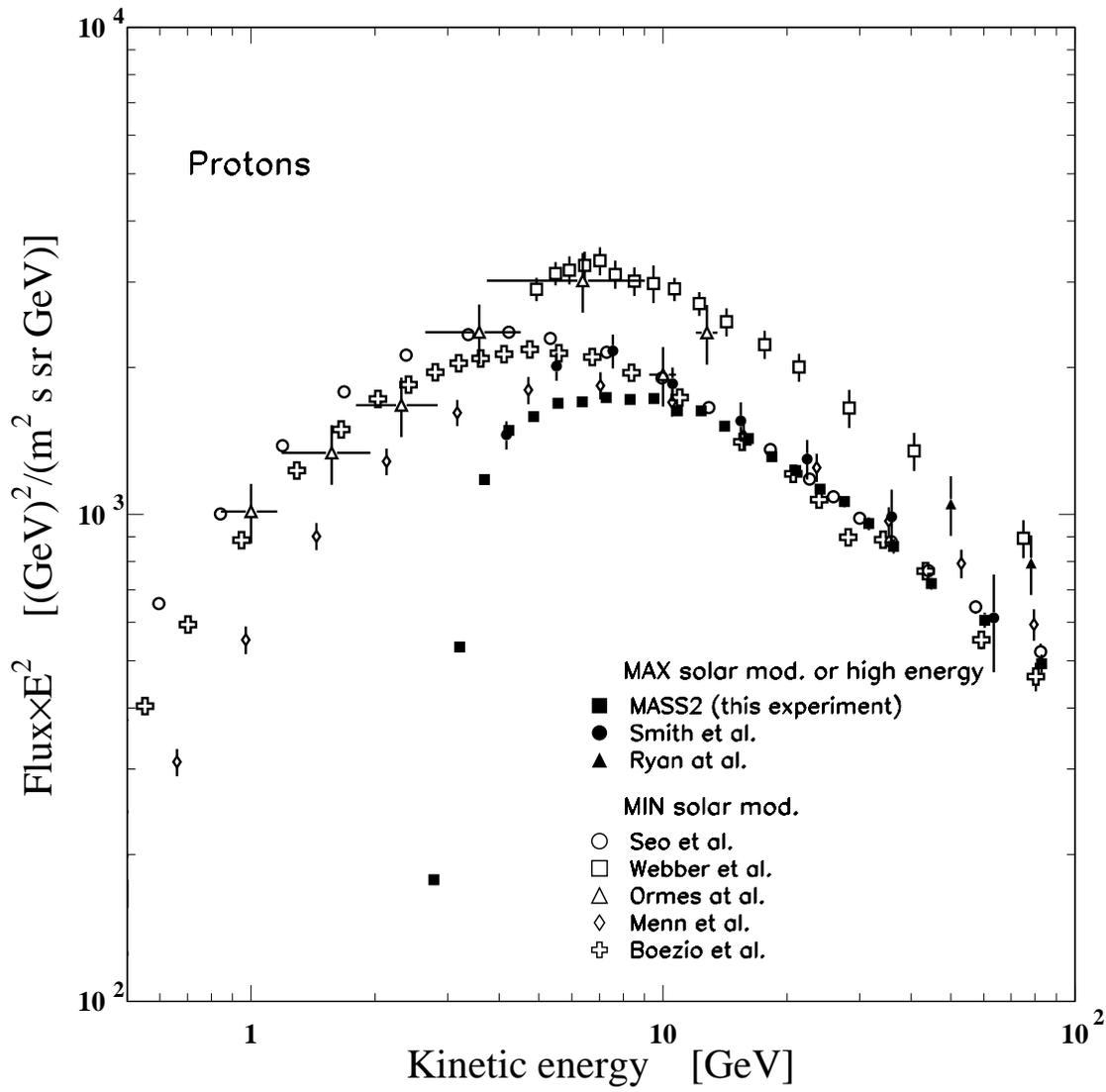,width=16cm}}
\caption{Comparison between the proton  flux measured in this
experiment and other data:
Smith {\it et al.} \protect\cite{b:smith},
Ryan {\it et al.} \protect\cite{b:ryan},
Seo {\it et al.} \protect\cite{b:seo},
Webber {\it et al.} \protect\cite{b:webber},
Ormes {\it et al.} \protect\cite{b:ormes},
Menn {\it et al.} \protect\cite{b:menn},
Boezio {\it et al.} \protect\cite{b:boezio}.
\label{f:proton}}  
\end{figure}

\begin{figure}
\mbox{\epsfig{file=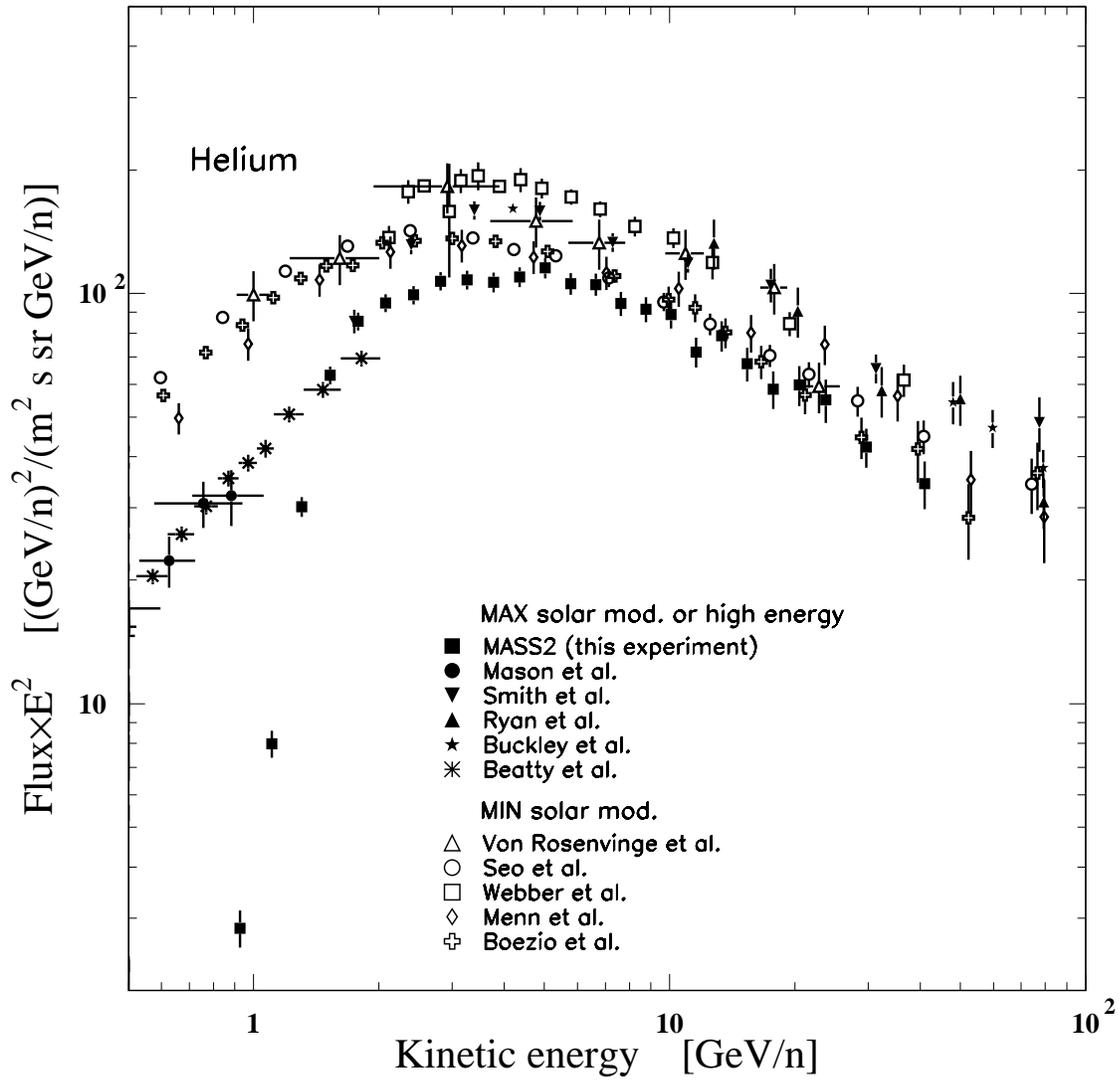,width=16cm}}
\caption{Comparison between the helium  flux 
measured in this experiment and other data:
Smith {\it et al.} \protect\cite{b:smith},
Ryan {\it et al.} \protect\cite{b:ryan},
Seo {\it et al.} \protect\cite{b:seo},
Webber {\it et al.} \protect\cite{b:webber},
Menn {\it et al.} \protect\cite{b:menn},
Boezio {\it et al.} \protect\cite{b:boezio}, 
Mason {\it et al.} \protect\cite{b:meson},
Von Rosenvinge {\it et al.} \protect\cite{b:vonrosenvinge},
Buckley {\it et al.} \protect\cite{buckley},
Beatty {\it et al.} \protect\cite{beatty}.
\label{f:helium}}  
\end{figure}

\begin{figure}
\mbox{\epsfig{file=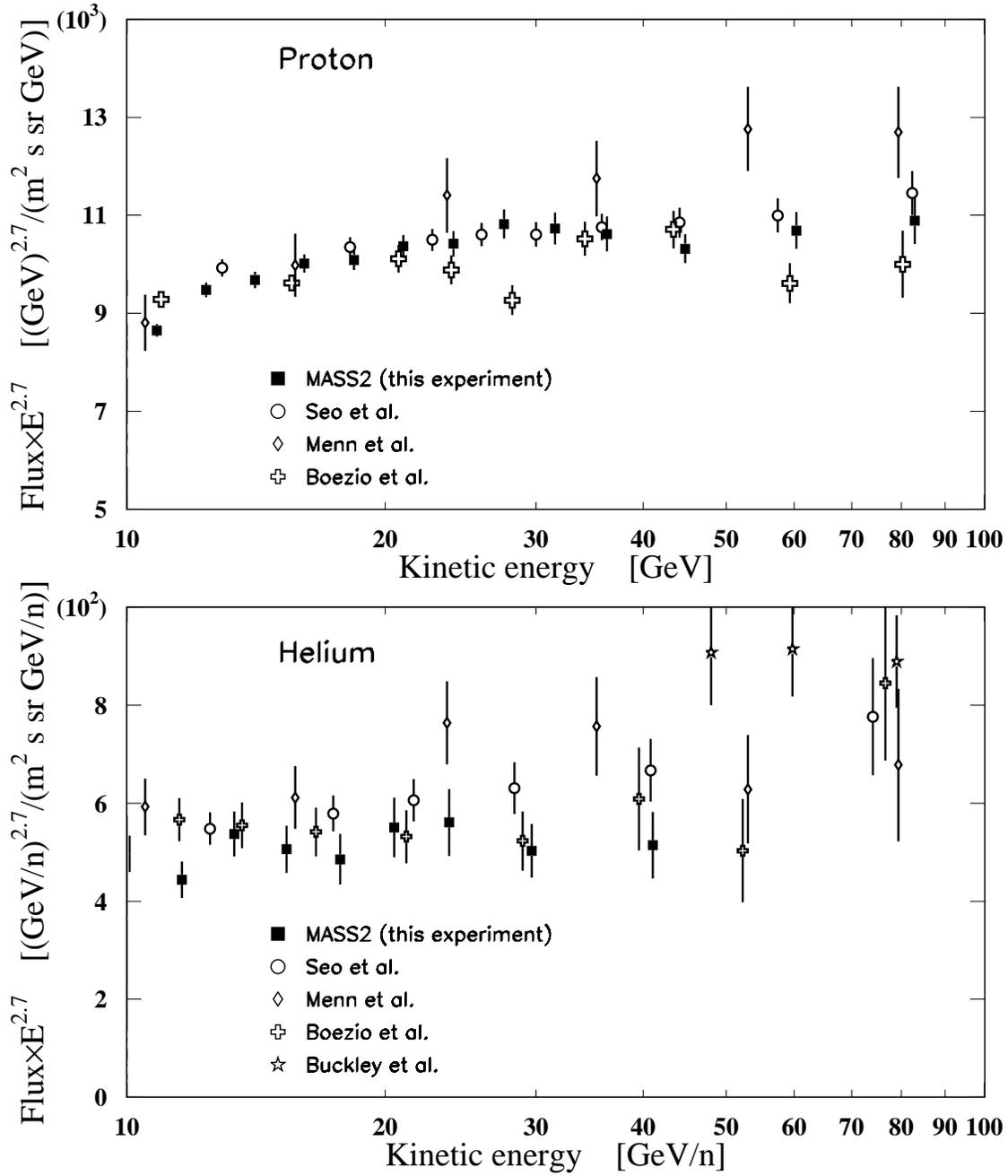,width=16cm}}
\caption{Comparison of the most recent high-energy
measurements of proton and helium fluxes:
Seo {\it et al.} \protect\cite{b:seo},
Menn {\it et al.} \protect\cite{b:menn},
Boezio {\it et al.} \protect\cite{b:boezio},
Buckley {\it et al.} \protect\cite{buckley}.
\label{f:recenti}}  
\end{figure}

The comparison of the results from this experiment 
with data from other experiments is shown in
Fig.~\ref{f:proton} for protons and in Fig.~\ref{f:helium} for helium. 
In general, 
it seems that there are several
inconsistencies among the different experiments. 
Such  discrepancies cannot be ascribed  completely
to the solar modulation effect, since they are noticed even 
at high energies where the solar modulation effect is very small.
If we compare only the most recent data, 
as shown in Fig.~\ref{f:recenti} for energies above 10 GeV/n, 
we see that the discrepancies are reduced. However, the differences 
between different data in some cases are of the order of 20--30\%,
considerably larger than the estimated errors. 
It is difficult
to establish a priori what systematics affect the different
experiments. Therefore, in order to avoid the effect of such systematic
errors in the  comparison between atmospheric and primary cosmic 
ray fluxes, the approach  proposed in this paper is to use the same 
apparatus to measure both
the atmospheric muons and their parent primary particle
fluxes. 

\section{ Conclusions}
We have reported on  simultaneous measurements 
of atmospheric muons and of primary cosmic rays taken with 
the same apparatus in a balloon experiment. 
The muon measurements cover the atmospheric depth range 
between 5 and 886 g/cm$^2$.
Negative muon  spectra were measured  in the 
momentum range 0.3--40 GeV/$c$, while positive muons 
between 0.3 and 1.5 GeV/$c$.
The proton and helium measurements were carried out
 at 5.8 g/cm$^2$, in the 3--100 GV rigidity range.
Corrections were applied in order to calculate the expected primary
fluxes at the top of the atmosphere.
The geomagnetic transmission function at
mid--latitude has been determined.
The data analysis procedures for primary nuclei and muon fluxes
were similar. Nevertheless, some differences in the selection
criteria for different particles were used. For this reason we can estimate a
normalization uncertainty of 1\% between proton and negative muon fluxes, 
and of 2\% between proton and positive muon fluxes.
The availability of results of muons and primaries taken  
with the same detector in the same experiment
may help decrease the uncertainties in the 
atmospheric neutrino
calculations.

\acknowledgments
We acknowledge very useful 
discussions with T. Stanev,  T. K. Gaisser 
and also with V. A. Naumov.
We thank the National Scientific Balloon Facility 
(Palestine,
Texas), which operated the flight.
This work was supported by NASA Grant~NAG-110, 
DARA and DFG, Germany,
the Istituto Nazionale
di Fisica Nucleare, Italy, and the Agenzia Spaziale Italiana, as
part of the research activities of the WIZARD collaboration.
A special thank to our technical support staff from NMSU and INFN.


\newpage

\clearpage

\clearpage

\clearpage

\clearpage

\clearpage

\clearpage

\clearpage

\clearpage

\clearpage

\clearpage

\clearpage

\clearpage

\clearpage

\clearpage

\clearpage

\clearpage

\clearpage

\narrowtext
\begin{table}
\caption{Selection criteria for track reconstruction
\label{ttracce}}
\begin{tabular}{cl} 
\multicolumn{1}{c}{Test} & \multicolumn{1}{c}{Description} \\ 
\tableline
1 & at least 11 signals in the spectrometer along the  
$x$-view\tablenotemark[1]  \\
2 & at least 8  signals in the spectrometer along the
$y$-view\tablenotemark[1]  \\
3 & not more than 3 multiple hits,\tablenotemark[2]
$x$-view\tablenotemark[1] \\
4 & not more than 3 multiple hits,\tablenotemark[2] 
$y$-view\tablenotemark[1] \\
5 & $\chi_{x}^{2}\leq8$  and $\chi_{y}^{2}\leq8$ \\
6 & the reconstructed deflection uncertainty $\sigma_{\eta}\leq0.03$
GV$^{-1}$ \\ 
7 & extrapolated track and positions in the scintillators\tablenotemark[3]  
consistent 
within 10~cm\\
8 & reconstructed track crossing the calorimeter \\
9 & extrapolation of the track not intersecting the lift bar \\
10 & $\beta$, as measured from the T.O.F., between 0 and 2 \\
\end{tabular}
\tablenotetext[1]{The $x$-view is the maximum curvature axis, the $y$-view
the perpendicular direction. }
\tablenotetext[2]{Multiple hits were defined as signals in the drift 
chambers at a distance larger than 4~cm from the reconstructed track. }
\tablenotetext[3]{The crossing positions 
in the scintillators were estimated by 
using the time signals from the T.O.F.\ system.}
\end{table}

\narrowtext
\begin{table}
\caption{Selection criteria for muons
\label{tmuoni}}
\begin{tabular}{cl} 
\multicolumn{1}{c}{Test} & \multicolumn{1}{c}{Description} \\ 
\tableline
1 & pulse height from the top scintillators\tablenotemark[1]
 between 0.7 and 1.8 $I_0$ \\
2 & number of $x$-view hits in the calorimeter between 2 and 10 \\
3 & number of $y$-view hits in the calorimeter between 2 and 14 \\
4 & not more than 1 multiple calorimeter hits along each view \\
5 & Cherenkov signal less than the equivalent to 3 
photoelectrons at less than 0.8~GV \\
6 & squared mass\tablenotemark[2] $m^2 
\leq 0.5$ GeV$^2$ in the rigidity range 0.65--1.25 GV \\
  &  squared mass\tablenotemark[2] $m^2 
\leq 0.3$ GeV$^2$ in the rigidity range 1.25--1.5 GV \\
\end{tabular}
\tablenotetext[1]{as determined from~(\protect\ref{mprotscint}).}
\tablenotetext[2]{as from equation~(\protect\ref{mmassa}).} 
\end{table}

\mediumtext
\begin{table}
\caption{Sources for muon background
\label{tcontaminazione}}
\begin{tabular}{ccc} 
source & \multicolumn{1}{c}{Rejection criteria} & Residual contamination \\ 
\tableline
albedo\tablenotemark[1]  & T.O.F.\ measurement & none \\
low-energy electrons & Cherenkov test 
& $\lesssim$1\% below 0.5 GV \\
and positrons\tablenotemark[2] &      &       \\
spillover  protons\tablenotemark[1]  & no correction & negligible \\
atmospheric mesons\tablenotemark[2]   & no correction 
&$\lesssim$1-2\% almost everywhere \\
 	&	&  for $\lesssim$10 GeV/$c$ pions; negligible for kaons \\
locally-produced 
& track reconstruction &
$\lesssim$5\% above 50 g/cm$^2$ 
below 1 GeV/$c$; \\
mesons\tablenotemark[2]  &
requirements & 
 negligible above 1 GeV/$c$ \\
low-energy protons\tablenotemark[3] & T.O.F.\ measurement &
$\lesssim$1\%
\\
\end{tabular}
\tablenotetext[1]{For $\mu^-$.}
\tablenotetext[2]{Both for $\mu^+$ and $\mu^-$.}
\tablenotetext[3]{For $\mu^+$.}
\end{table}

\begin{table}
\caption{Flux growth curve results for 0.3--40 GeV/$c$ negative 
and 0.3--1.5 GeV/$c$ positive muons.
Results are given for the following momentum intervals:
I 0.3--0.9 GeV/$c$, II 0.9--1.5 GeV/$c$, III 1.5--2.5 GeV/$c$, 
IV 2.5--4 GeV/$c$, V 4--40 GeV/$c$.
The symbols APD and FAD stand respectively for Average Payload Depth and
Flux-weighted Average Depth in the momentum bin. 
In the latter case, the best fit curve of type~(\protect\ref{mprofilo})
was used. 
The units of flux are \mbox{particles/(cm$^2$ s sr GeV/$c$)}.
\label{tcurve}}
\begin{tabular}{clcccc} 
\multicolumn{2}{c}{Depth Interval} & A & B & C & D \\ \tableline
\multicolumn{2}{l}{Duration (s)} & 820 &  530 &  350 & 390 \\ 
\multicolumn{2}{l}{Live-time fraction} & 0.920 & 0.818 & 0.725 & 0.648 \\ 
\multicolumn{2}{l}{Initial depth (g/cm$^2$)} & 886 & 514 & 358 & 272 \\ 
\multicolumn{2}{l}{Final depth (g/cm$^2$)} & 514 & 358 & 272 & 197 \\ 
\multicolumn{2}{l}{APD 
 (g/cm$^2$)} & 680 & 434 & 316 & 233 \\ 
\cline{1-2}\\
I & $\mu^-$ Flux 
 & 2.59$\pm$0.28$\cdot 10^{-3}$ &  6.49$\pm$0.59$\cdot 
10^{-3}$ &  9.79$\pm$0.95$\cdot 10^{-3}$ &  1.03$\pm$0.10$\cdot 10^{-2}$   
\\
& FAD 
 (g/cm$^2$) & 631.6 &  421.3 &  309.1 &  228.7 \\
II & $\mu^-$ Flux 
 & 2.25$\pm$0.24$\cdot 10^{-3}$ &  4.21$\pm$0.43$\cdot 
10^{-3}$ &  5.87$\pm$0.66$\cdot 10^{-3}$ &  6.33$\pm$0.69$\cdot 10^{-3}$   
\\
& FAD 
 (g/cm$^2$) & 637.4 &  422.6 &  309.5 &  229.0 \\
III & $\mu^-$ Flux 
 & 1.24$\pm$0.14$\cdot 10^{-3}$ &  2.13$\pm$0.23$\cdot 
10^{-3}$ &  2.70$\pm$0.34$\cdot 10^{-3}$ &  3.44$\pm$0.39$\cdot 10^{-3}$   
\\
& FAD 
 (g/cm$^2$) & 639.5 &  423.0 &  309.7 &  229.1 \\
IV & $\mu^-$ Flux 
 & 5.30$\pm$0.71$\cdot 10^{-4}$ & 8.89$\pm$1.22$\cdot 
10^{-4}$ &  1.10$\pm$0.18$\cdot 10^{-3}$ &  1.24$\pm$0.19$\cdot 10^{-3}$   
\\
& FAD 
 (g/cm$^2$) & 645.0 &  424.2 &  310.0 &  229.4 \\
V & $\mu^-$ Flux 
 & 4.25$\pm$0.41$\cdot 10^{-5}$ &  5.36$\pm$0.61$\cdot 
10^{-5}$ &  6.49$\pm$0.87$\cdot 10^{-5}$ &  5.82$\pm$0.83$\cdot 10^{-5}$   
\\
& FAD 
 (g/cm$^2$) & 650.3 &  425.3 &  310.4 &  229.6 \\
\cline{1-2}\\
I & $\mu^+$ Flux 
 & 2.84$\pm$0.29$\cdot 10^{-3}$ &  6.99$\pm$0.62$\cdot 
10^{-3}$ &  9.30$\pm$0.92$\cdot 10^{-3}$ &  1.30$\pm$0.11$\cdot 10^{-2}$   
\\
& FAD 
 (g/cm$^2$) & 630.5 &  421.1 &  309.0 &  228.6 \\
II & $\mu^+$ Flux 
 & 2.92$\pm$0.28$\cdot 10^{-3}$ &  4.85$\pm$0.47$\cdot 
10^{-3}$ &  6.80$\pm$0.72$\cdot 10^{-3}$ &  7.79$\pm$0.77$\cdot 10^{-3}$   
\\
& FAD 
 (g/cm$^2$) & 637.2 &  422.6 &  309.5 &  229.0 \\
\end{tabular}
\end{table}

\addtocounter{table}{-1}
\widetext
\begin{table}
\caption{(continued)}
\begin{tabular}{clcccc} 
\multicolumn{2}{c}{Depth Interval} & E & F & G & H \\ \tableline
\multicolumn{2}{l}{Duration (s)} & 510 &  670 &  670 & 780 \\ 
\multicolumn{2}{l}{Live-time fraction} & 0.594 & 0.597 & 0.593 & 0.602 \\ 
\multicolumn{2}{l}{Initial depth (g/cm$^2$)} & 197 & 134 & 95 & 69 \\ 
\multicolumn{2}{l}{Final depth (g/cm$^2$)} & 134 & 95 & 69 & 48 \\ 
\multicolumn{2}{l}{APD 
 (g/cm$^2$)} & 163 & 112 & 81 & 59 \\ 
\cline{1-2}\\
I & $\mu^-$ Flux 
& 1.16$\pm$0.10$\cdot 10^{-2}$ &  9.93$\pm$0.78$\cdot 
10^{-3}$ &  8.85$\pm$0.74$\cdot 10^{-3}$ &  7.64$\pm$0.63$\cdot 10^{-3}$   
\\
& FAD 
(g/cm$^2$) & 160.4 &  110.5 &  80.4 &  58.2 \\
II & $\mu^-$ Flux 
& 6.09$\pm$0.64$\cdot 10^{-3}$ &  6.11$\pm$0.54$\cdot 
10^{-3}$ &  5.35$\pm$0.51$\cdot 10^{-3}$ &  4.17$\pm$0.41$\cdot 10^{-3}$   
\\
& FAD 
(g/cm$^2$) & 160.6 &  110.6  &  80.4 &  58.2 \\
III & $\mu^-$ Flux 
& 3.23$\pm$0.34$\cdot 10^{-3}$ &  2.80$\pm$0.28$\cdot 
10^{-3}$ &  2.57$\pm$0.27$\cdot 10^{-3}$ &  1.90$\pm$0.21$\cdot 10^{-3}$   
\\
& FAD 
(g/cm$^2$) & 160.7 &  110.6 &  80.5 &  58.2 \\
IV & $\mu^-$ Flux 
& 1.32$\pm$0.18$\cdot 10^{-3}$ &  9.29$\pm$1.30$\cdot 
10^{-4}$ &  8.45$\pm$1.24$\cdot 10^{-4}$ &  6.54$\pm$1.00$\cdot 10^{-4}$   
\\
& FAD 
(g/cm$^2$) & 160.9 &  110.7 &  80.5 &  58.2 \\
V & $\mu^-$ Flux 
& 4.95$\pm$0.70$\cdot 10^{-5}$ &  4.71$\pm$0.59$\cdot 
10^{-5}$ &  3.78$\pm$0.53$\cdot 10^{-5}$ &  3.01$\pm$0.44$\cdot 10^{-5}$   
\\
& FAD 
(g/cm$^2$) & 161.1  &  110.8 &  80.5 &  58.2 \\
\cline{1-2}\\
I & $\mu^+$ Flux 
& 1.29$\pm$0.10$\cdot 10^{-2}$ &  1.16$\pm$0.08$\cdot 
10^{-2}$ &  9.71$\pm$0.77$\cdot 10^{-3}$ &  8.56$\pm$0.66$\cdot 10^{-3}$   
\\
& FAD 
(g/cm$^2$) & 160.4 &  110.5 &  80.4 &  58.2 \\
II & $\mu^+$ Flux 
& 8.46$\pm$0.69$\cdot 10^{-3}$ &  7.54$\pm$0.61$\cdot 
10^{-3}$ &  5.54$\pm$0.52$\cdot 10^{-3}$ &  4.41$\pm$0.43$\cdot 10^{-3}$   
\\
& FAD 
(g/cm$^2$) & 160.6 &  110.6 &  80.4 &  58.2 \\
\end{tabular}
\end{table}

\addtocounter{table}{-1}
\widetext
\begin{table}
\caption{(continued)}
\begin{tabular}{clcc} 
\multicolumn{2}{c}{Depth Interval} & I  & J \\ \tableline
\multicolumn{2}{l}{Duration (s)} &  1100 & 3590 \\ 
\multicolumn{2}{l}{Live-time fraction} & 0.619 & 0.646 \\ 
\multicolumn{2}{l}{Initial depth (g/cm$^2$)} & 48  & 27 \\ 
\multicolumn{2}{l}{Final depth (g/cm$^2$)} & 27 & 5 \\ 
\multicolumn{2}{l}{APD 
 (g/cm$^2$)} & 37 & 13 \\ 
\cline{1-2}\\
I & $\mu^-$ Flux 
& 4.78$\pm$0.41$\cdot 10^{-3}$ &  2.00$\pm$0.14$\cdot 
10^{-3}$ 
\\
& FAD 
(g/cm$^2$) & 37.4 &  15.3 \\
II & $\mu^-$ Flux 
& 2.44$\pm$0.26$\cdot 10^{-3}$ &  1.02$\pm$0.09$\cdot 
10^{-3}$ 
\\
& FAD 
(g/cm$^2$) & 37.4 &  15.3 \\
III & $\mu^-$ Flux 
& 1.29$\pm$0.14$\cdot 10^{-3}$ &  5.11$\pm$0.50$\cdot 
10^{-4}$ 
\\
& FAD 
(g/cm$^2$) & 37.4 &  15.3 \\
IV & $\mu^-$ Flux 
& 4.20$\pm$0.67$\cdot 10^{-4}$ &  1.60$\pm$0.22$\cdot 
10^{-4}$ 
\\
& FAD 
(g/cm$^2$) & 37.4 &  15.4 \\
V & $\mu^-$ Flux 
& 2.50$\pm$0.33$\cdot 10^{-5}$ &  9.63$\pm$1.12$\cdot 
10^{-6}$ 
\\
& FAD 
(g/cm$^2$) & 37.4 &  15.4 \\
\cline{1-2}\\
I & $\mu^+$ Flux 
& 5.57$\pm$0.45$\cdot 10^{-3}$ &  2.38$\pm$0.16$\cdot 
10^{-3}$ 
\\
& FAD 
(g/cm$^2$) & 37.4 &  15.3 \\
II & $\mu^+$ Flux 
& 3.08$\pm$0.30$\cdot 10^{-3}$ &  1.41$\pm$0.11$\cdot 
10^{-3}$ 
\\
& FAD 
(g/cm$^2$) & 37.4 &  15.3 \\
\end{tabular}
\end{table}

\begin{table}
\caption{Negative muon spectra in different depth intervals. 
The results are given 
for the following momentum bins:
I 0.3--0.465 GeV/$c$, II 0.465--0.65 GeV/$c$, III 0.65--0.8 GeV/$c$, 
IV 0.8--1 GeV/$c$, V 1--1.25 GeV/$c$,
VI 1.25--1.5 GeV/$c$, VII 1.5--2 GeV/$c$, 
VIII 2--3 GeV/$c$, IX 3--8 GeV/$c$, X 8--40 GeV/$c$.
The symbols APD and FAD stand respectively for Average Payload Depth and
Flux-weighted Average Depth in the momentum bin. 
In the latter case, the best fit curve of type~(\protect\ref{mprofilo})
was used. 
The units of flux are \mbox{particles/(cm$^2$ s sr GeV/$c$)}.
\label{tspettri}}
\begin{tabular}{clccccc} 
\multicolumn{2}{c}{Depth Interval} & A & B & C & D  & E \\ \tableline
\multicolumn{2}{l}{Duration (s)} & 540 &  700 &  510 & 1190 & 1230 \\ 
\multicolumn{2}{l}{Live-time fraction} & 0.627 & 
0.594 & 0.592 & 0.600 & 0.620 \\ 
\multicolumn{2}{l}{Initial depth (g/cm$^2$)} &
255 & 164 & 106 & 83 & 48 \\
\multicolumn{2}{l}{Final depth (g/cm$^2$)} 
& 164 & 106 & 83 & 48 & 25 \\ 
\multicolumn{2}{l}{APD 
 (g/cm$^2$)} & 206 & 131 & 94 &  65 & 36 \\ 
\multicolumn{2}{l}{FAD 
 (g/cm$^2$)} & 202 & 130 & 93 &  65 & 36  \\
\cline{1-2}\\
I & $\mu^-$ Flux 
 & 1.21$\pm$0.20$\cdot 10^{-2}$ &  1.48$\pm$0.20$\cdot 
10^{-2}$ &  1.16$\pm$0.20$\cdot 10^{-2}$ &  1.06$\pm$0.12$\cdot 10^{-2}$   
&  5.58$\pm$0.88$\cdot 10^{-3}$   \\
II & $\mu^-$ Flux 
 & 1.06$\pm$0.15$\cdot 10^{-2}$ &  1.22$\pm$0.14$\cdot 
10^{-2}$ &  8.58$\pm$1.42$\cdot 10^{-3}$ &  7.96$\pm$0.90$\cdot 10^{-3}$   
&  5.13$\pm$0.69$\cdot 10^{-3}$   \\
III & $\mu^-$ Flux 
 & 9.00$\pm$1.45$\cdot 10^{-3}$ &  9.46$\pm$1.35$\cdot 
10^{-3}$ &  7.78$\pm$1.43$\cdot 10^{-3}$ &  6.63$\pm$0.86$\cdot 10^{-3}$   
&  4.18$\pm$0.66$\cdot 10^{-3}$   \\
IV & $\mu^-$ Flux 
 & 8.41$\pm$1.20$\cdot 10^{-3}$ &  7.82$\pm$1.04$\cdot 
10^{-3}$ &  6.98$\pm$1.15$\cdot 10^{-3}$ &  6.47$\pm$0.73$\cdot 10^{-3}$   
&  3.29$\pm$0.50$\cdot 10^{-3}$   \\
V & $\mu^-$ Flux 
 & 6.09$\pm$0.90$\cdot 10^{-3}$ &  6.91$\pm$0.87$\cdot 
10^{-3}$ &  6.39$\pm$0.98$\cdot 10^{-3}$ &  4.90$\pm$0.56$\cdot 10^{-3}$   
&  2.76$\pm$0.40$\cdot 10^{-3}$   \\
VI & $\mu^-$ Flux 
 & 4.83$\pm$0.80$\cdot 10^{-3}$ &  4.90$\pm$0.73$\cdot 
10^{-3}$ &  4.69$\pm$0.83$\cdot 10^{-3}$ &  3.29$\pm$0.45$\cdot 10^{-3}$   
&  1.86$\pm$0.33$\cdot 10^{-3}$   \\
VII & $\mu^-$ Flux 
 & 4.13$\pm$0.52$\cdot 10^{-3}$ &  3.31$\pm$0.42$\cdot 
10^{-3}$ &  3.55$\pm$0.51$\cdot 10^{-3}$ &  2.73$\pm$0.29$\cdot 10^{-3}$   
&  1.60$\pm$0.22$\cdot 10^{-3}$   \\
VIII & $\mu^-$ Flux 
 & 2.58$\pm$0.29$\cdot 10^{-3}$ &  2.18$\pm$0.24$\cdot 
10^{-3}$ &  1.43$\pm$0.23$\cdot 10^{-3}$ &  1.30$\pm$0.14$\cdot 10^{-3}$   
&  7.21$\pm$1.02$\cdot 10^{-4}$   \\
IX & $\mu^-$ Flux 
 & 4.79$\pm$0.56$\cdot 10^{-4}$ &  2.97$\pm$0.39$\cdot 
10^{-4}$ &  4.02$\pm$0.54$\cdot 10^{-4}$ &  2.39$\pm$0.27$\cdot 10^{-4}$   
&  1.96$\pm$0.24$\cdot 10^{-4}$   \\
X & $\mu^-$ Flux 
 & 2.44$\pm$0.49$\cdot 10^{-5}$ &  2.30$\pm$0.43$\cdot 
10^{-5}$ &  1.64$\pm$0.42$\cdot 10^{-5}$ &  1.43$\pm$0.26$\cdot 10^{-5}$   
&  8.66$\pm$1.94$\cdot 10^{-6}$   \\
\end{tabular}
\end{table}

\begin{table}
\caption{Proton flux at the top of the atmosphere. \label{t:pr_flux}}
\begin{tabular}{rlccc}
\multicolumn{2}{c}{Kin.\ energy range} 
& mean energy & flux & flux error \\ 
\multicolumn{2}{c}{(GeV)} & (GeV) & \multicolumn{2}{c}{(m$^2$ sr s GeV)$^{-1}$}
\\ \hline
   2.55 &   2.95 &  2.77 &   23.10 &   0.48   \\
   2.95 &   3.41 &  3.21 &   51.84 &   0.71  \\
   3.41 &   3.93 &  3.68 &   86.90 &   0.92  \\
   3.93 &   4.52 &  4.22 &   83.46 &   0.85  \\
   4.52 &   5.19 &  4.85 &   67.66 &   0.72  \\
   5.19 &   5.95 &  5.56 &   54.80 &   0.60  \\
   5.95 &   6.81 &  6.36 &   42.04 &   0.48  \\
   6.81 &   7.78 &  7.28 &   32.83 &   0.40  \\
   7.78 &   8.89 &  8.31 &   24.90 &   0.32  \\
   8.89 &   10.1 &  9.49 &   19.18 &   0.26  \\
   10.1 &   11.6 &  10.8 &   13.90 &   0.21  \\
   11.6 &   13.2 &  12.3 &   10.66 &   0.17  \\
   13.2 &   15.1 &  14.1 &   7.63  &   0.13  \\
   15.1 &   17.2 &  16.1 &   5.52  &   0.10  \\
   17.2 &   19.7 &  18.3 &   3.89  &   0.08  \\
   19.7 &   22.5 &  21.0 &   2.79  &   0.06  \\
   22.5 &   25.7 &  24.0 &   1.95  &   0.05  \\
   25.7 &   29.5 &  27.5 &   1.40  &   0.04  \\
   29.5 &   33.9 &  31.6 &   0.961 &   0.029 \\
   33.9 &   39.0 &  36.3 &   0.652 &   0.022 \\
   39.0 &   52.2 &  44.8 &   0.360 &   0.010 \\
   52.2 &   71.0 &  60.3 &   0.167 &   0.006 \\
   71.0 &   99.1 &  83.0 &   0.072 &   0.003 \\
\end{tabular}
\end{table}


\begin{table}
\caption{Helium flux at the top of the atmosphere. \label{t:he_flux}}
\begin{tabular}{rlccc}
\multicolumn{2}{c}{Kin.\ energy range} 
& mean energy & flux & flux error \\ 
\multicolumn{2}{c}{(GeV/n)} & (GeV/n) & \multicolumn{2}{c}{(m$^2$ sr s
GeV/n)$^{-1}$} \\ \hline

 .841 &  1.00 &  .927 &  3.30 &  .34  \\
 1.00 &  1.19 &  1.11 &  6.53 &  .49  \\
 1.19 &  1.40 &  1.30 &  17.7 &  1.0  \\
 1.40 &  1.64 &  1.52 &  27.2 &  1.4  \\
 1.64 &  1.92 &  1.78 &  27.0 &  1.3  \\
 1.92 &  2.24 &  2.08 &  21.9 &  1.1  \\
 2.24 &  2.62 &  2.42 &  16.9 &  0.9  \\
 2.62 &  3.03 &  2.82 &  13.5 &  0.7  \\
 3.03 &  3.51 &  3.26 &  10.1 &  0.5  \\
 3.51 &  4.06 &  3.77 &  7.48 &  0.41  \\
 4.06 &  4.68 &  4.36 &  5.78 &  0.32  \\
 4.68 &  5.39 &  5.02 &  4.58 &  0.26  \\
 5.39 &  6.20 &  5.77 &  3.17 &  0.19  \\
 6.20 &  7.13 &  6.64 &  2.38 &  0.15  \\
 7.13 &  8.20 &  7.63 &  1.62 &  0.11  \\
 8.20 &  9.42 &  8.77 &  1.19 &  0.08  \\
 9.42 &  10.8 &  10.1 &  0.875 & 0.065  \\
 10.8 &  12.4 &  11.6 &  0.536 & 0.045  \\
 12.4 &  14.3 &  13.3 &  0.444 & 0.038  \\
 14.3 &  16.5 &  15.3 &  0.286 & 0.027  \\
 16.5 &  19.1 &  17.7 &  0.186 & 0.020  \\
 19.1 &  22.1 &  20.5 &  0.142 & 0.016  \\
 22.1 &  25.7 &  23.7 &  0.098 & 0.011  \\
 25.7 &  35.0 &  29.7 &  0.048 & 0.005  \\
 35.0 &  49.1 &  41.0 &  0.020 & 0.0027  \\
\end{tabular}
\end{table}

\clearpage
\end{document}